\documentclass[twocolumn]{aastex631}

\usepackage{CJKutf8}
\usepackage{natbib}
\usepackage{multirow}
\usepackage{graphicx}
\usepackage{subfigure}
\usepackage{float}
\usepackage{footmisc}
\usepackage{amsmath}

\newcommand{\siiii}{Si {\sc iii} $\lambda$1206}
\newcommand{\siiv}{Si {\sc iv} $\lambda$$\lambda$1394, 1403}
\newcommand{\siiva}{Si {\sc iv} $\lambda$1394}
\newcommand{\siivb}{Si {\sc iv} $\lambda$1394}

\newcommand{\oiitext}{[O {\sc ii}]}
\newcommand{\oiiitext}{[O {\sc iii}]}

\newcommand{\civ}{C {\sc iv} $\lambda$$\lambda$1548, 1550}
\newcommand{\civa}{C {\sc iv} $\lambda$1548}
\newcommand{\civb}{C {\sc iv} $\lambda$1550}
\newcommand{\cii}{C {\sc ii} $\lambda$1335}
\newcommand{\ciiex}{C {\sc ii}* $\lambda$1336}
\newcommand{\heii}{He {\sc ii} $\lambda$1640}

\newcommand{\oivb}{O {\sc iv}] $\lambda$1404}

\newcommand{\sivi}{[Si {\sc vi}] 1.9630 \mum}

\newcommand{\oivtext}{O {\sc iv}]}

\newcommand{\civtext}{C~{\sc iv}}
\newcommand{\ciitext}{C~{\sc ii}}
\newcommand{\ciiext}{C~{\sc ii*}}
\newcommand{\heiitext}{He~{\sc ii}}

\newcommand{\siivtext}{Si~{\sc iv}}
\newcommand{\siiitext}{Si~{\sc ii}}

\newcommand{\msun}{\ensuremath{\mathrm{M}_{\odot}}}

\newcommand{\msunyr}{$M_{\sun}\ yr^{-1}$}

\newcommand{\chitwo}{$\chi^{2}\ $}

\newcommand{\kms}{km\ s$^{-1}$}

\newcommand{\nel}{$n_{e}$}
\newcommand{\NH}{$N_{H}$}

\def\cm3{~cm$^{-3}$}

\newcommand{\vesc}{$v_{esc}$}
\newcommand{\vcirc}{$v_{circ}$}

\newcommand{\dedt}{$\dot{E}_{out}$}
\newcommand{\dedtlagn}{$\dot{E}_{out}/L_{AGN}$}

\newcommand{\mum}{\ifmmode{\rm \mu m}\else{$\mu$m}\fi}

\newcommand{\vwu}{{$v_{50}$}}

\newcommand{\oiiab}{[O~{\sc ii}] $\lambda$$\lambda$3726,3729}
\newcommand{\oiii}{[O~{\sc iii}] $\lambda$5007}

\newcommand{\lagn}{L$_{AGN}$}

\newcommand{\aox}{$\alpha_{OX}$}

\newcommand{\heiiopt}{He~{\sc ii} $\lambda$4686}

\newcommand{\siiic}{Si~{\sc ii} $\lambda$1260}

\newcommand{\siiie}{Si~{\sc ii} $\lambda$1526}
\newcommand{\siiietext}{Si~{\sc ii}}

\newcommand{\siiiabcde}{Si~{\sc ii} 1190, 1193, 1260, 1304, 1526}

\newcommand{\oiuv}{O {\sc i} $\lambda$1302}
\newcommand{\vcir}{$v_{circ}$}

\begin{document}

\title{Fast Outflows and Luminous He II Emission in Dwarf Galaxies with AGN}

\author[0000-0003-3762-7344]{Weizhe Liu \begin{CJK}{UTF8}{gbsn}(刘伟哲)\end{CJK}}
\affiliation{Steward Observatory, University of Arizona, 933 N. Cherry Ave., Tucson, AZ 85721, USA}
\correspondingauthor{Weizhe Liu}
\email{oscarlwz@gmail.com}

\author[0000-0002-3158-6820]{Sylvain Veilleux} 
\affiliation{Department of Astronomy, University of Maryland, College Park, MD 20742, USA}
\affiliation{Joint Space-Science
  Institute, University of Maryland, College Park, MD 20742, USA}

\author[0000-0003-4693-6157]{Gabriela Canalizo} 
\affiliation{Department of Physics and Astronomy University of California Riverside, 900 University Avenue, CA 92521, USA}

\author{Todd M. Tripp}
\affiliation{Department of Astronomy, University of Massachussetts, Amherst, MA 01003, USA}

\author[0000-0002-1608-7564]{David S. N. Rupke}
\affiliation{Department of Physics, Rhodes College, Memphis, TN 38112, USA}

\author[0000-0001-7578-2412]{Archana Aravindan}
\affiliation{Department of Physics and Astronomy, University of California, Riverside, 900 University Ave, Riverside CA 92521, USA}

\author[0000-0002-4375-254X]{Thomas Bohn} 
\affiliation{Department of Physics and Astronomy University of California Riverside, 900 University Avenue, CA 92521, USA}

\author{Fred Hamann}
\affiliation{Department of Physics and Astronomy, University of California, Riverside, CA 92507, USA}

 \author[0000-0002-5253-9433]{Christina M. Manzano-King} 
 \affiliation{Department of Physics and Astronomy University of California Riverside, 900 University Avenue, CA 92521, USA}

\begin{abstract}
While stellar processes are believed to be the main source of feedback in dwarf galaxies, the accumulating discoveries of AGN in dwarf galaxies over recent years arouse the interest to also consider AGN feedback in them. Fast, AGN-driven outflows, a major mechanism of AGN feedback, have indeed been discovered in dwarf galaxies and may be powerful enough to provide feedback to their dwarf hosts. In this paper, we search for outflows traced by the blueshifted ultraviolet absorption features in three dwarf galaxies with AGN from the sample examined in our previous ground-based study. We confirm outflows traced by blueshifted absorption features in two objects and tentatively detect an outflow in the third object. In one object where the outflow is clearly detected in multiple species, photoionization modeling suggests that this outflow is located $\sim$0.5 kpc from the AGN, implying a galactic-scale impact. This outflow is much faster and possesses higher kinetic energy outflow rate than starburst-driven outflows in sources with similar star formation rates, and is likely energetic enough to provide negative feedback to its host galaxy as predicted by simulations. Much broader ($\sim$4000 \kms) absorption features are also discovered in this object which may have the same origin as that of broad absorption lines in quasars. Additionally, strong \heii\ emission is detected in both objects where the transition falls in the wavelength coverage, and is consistent with an AGN origin. In one of these two objects, blueshifted \heii\ emission line is clearly detected, likely tracing a highly-ionized AGN wind.
\end{abstract}

\section{Introduction} \label{1}

Stellar processes have long been considered the main source of negative feedback in dwarf galaxies \citep[e.g.][]{Larson1974,Veilleux2005,Heckman2017,Martin2018}. However, it is still debated whether such stellar feedback is effective enough to reproduce all the related properties of the dwarf galaxies we see today \citep[e.g.][]{Garrison-Kimmel2013,McQuinn2019}. Recent studies have revealed hundreds of active galactic nuclei (AGN) in dwarf galaxies through multi-wavelength observations \citep[see the recent reviews by][]{Greene2019,Reines2022}. AGN activity can release a tremendous amount of energy, and AGN feedback has been widely accepted as a critical mechanism to regulate the formation and evolution of massive galaxies \citep[e.g.,][]{DiMatteo2005,Booth2009,Fabian2012araa}. Therefore, it is interesting to also consider the possible impact of AGN feedback in dwarf galaxies. Moreover, low-z dwarf galaxies likely possess properties close to those of moderate AGN host galaxies in the early universe \citep[e.g.,][]{Matthee2023,Maiolino2023}. Such AGN feedback may also regulate the growth of (super)massive black hole seeds and first galaxies \citep[e.g.][]{Silk2017}.

Evidence of AGN feedback in dwarf galaxies has emerged in recent years. Specifically, hints of star formation quenching induced by AGN feedback \citep{Penny2018} and lower-than-expectation global HI content \citep{Bradford2018} have both been reported in dwarf galaxies with AGN. Similar evidence on such feedback is also seen in several latest observational studies \citep[][]{Davis2022,Yang2023,Kawamuro2023}. From the theoretical perspective, analytic analysis \citep[e.g.][]{Silk2017,Dashyan2018} and new simulations \citep[e.g.][]{Koudmani2019,Koudmani2021,Lanfranchi2021} have all pointed out the possibly significant effects of AGN feedback in dwarfs. In addition to the negative feedback discussed above that suppresses star formation, AGN-driven outflows may also trigger star formation activities within dwarf galaxies \citep{SchutteReines2022}.

Recently, a sample of 29 dwarf galaxies with AGN (stellar mass log(M$_{\star}$/\msun) $<$\ 10.0) was observed with Keck optical spectroscopy \citep{ManzanoKing2019,ManzanoKing2020} and NIR spectroscopy \citep{Bohn2021}, revealing fast outflows in 9 of them.  This sample was chosen from dwarf galaxies at z $<$ 0.05 in SDSS showing properties in the optical consistent with AGN activity \citep[i.e., broad Balmer lines, Seyfert-like line ratios on the BPT/VO87 \citep{bpt,Veilleux1987} diagrams and/or highly-ionized \heiitext\ line emission;][]{Reines2013,Moran2014}. Follow-up integral field spectroscopy (IFS) of 8 of the 29 dwarf galaxies confirmed the existence of rapid outflows in 7 of them \citep[][L20 hereafter]{Liu2020}.  These outflows are primarily driven by the AGN, and are powerful enough to provide feedback for their host galaxies in a way similar to outflows of more luminous AGN with massive hosts. Additionally, such outflows are more powerful than those observed in a matched sample of non-AGN star-forming dwarf galaxies with similar galaxy properties \citep{Aravindan2023}.

While the results from the optical emission lines are tantalizing, they only probe the relatively denser part of the outflows as the emission line strength is proportional to the electron density squared. Absorption lines, instead, depend on column densities linearly and are thus more sensitive probes of the full extent of outflows captured along the line of sight. Moreover, blueshifted absorption features are unambiguous signatures of the outward motion of gas located in front of the source of continuum radiation. Rest-frame far-ultraviolet (FUV) spectroscopy provides access to multiple strong absorption features spanning a broad range of ionization states, which is essential for finer measurements and/or tighter upper limits on the physical conditions of outflows (e.g., more robust ionization correction). 

Such FUV spectroscopy has been used extensively to probe ionized and neutral gas outflows triggered by AGN \citep[e.g.][]{Crenshaw2012} and/or star formation \citep{Heckman2015,Heckman2016}, in galaxies with starbursts \citep[e.g.][]{Martin2015} or recently-quenched star formation \citep[e.g.][]{Tripp2011}. To evaluate the AGN feedback via fast outflows in dwarf galaxies with the same criterion, a sample of dwarf galaxies with primarily AGN-driven outflows needs to be examined with the same scrutiny.

In this paper, we present the results from a pilot \textit{HST}/COS FUV spectroscopic study of three dwarf galaxies with evidence of AGN-driven outflows from L20. In Section \ref{052}, the data sets, the physical properties of the sources measured from the \textit{HST}/COS and ancillary data, and the data reduction procedures are described. The detection and characterization of absorption line-traced outflows and thus AGN feedback within the sample are presented in Section \ref{053}. The analysis of the strong \heii\ emission lines detected in these objects is described in Section \ref{054}. Throughout the paper, we assume a $\Lambda$CDM cosmology with $H_0$ = 69.3 km s$^{-1}$ Mpc$^{-1}$, $\Omega_{\rm m}$ = 0.287, and $\Omega_{\rm \Lambda} = 0.713$ \citep{wmap2013}.

\section{HST and Ancillary Data} \label{052}

\subsection{HST/COS Observations and Data Reduction} \label{0521}

We observed the three FUV-brightest dwarf galaxies with AGN-driven outflows from the sample studied in L20 (see Table \ref{tab:targets} for object names and their short names). The targets were observed with \textit{HST}/COS through cycle 27 program PID 15915 (PI: W. Liu). The corresponding observations can be accessed via \dataset[10.17909/9rjd-2819]{http://dx.doi.org/10.17909/9rjd-2819}. The spectra of our targets were obtained in TIME-TAG mode through the Primary Science Aperture (PSA) using the medium resolution FUV grating, G160M. Four focal plane offset positions were adopted to reduce the impact of fixed-pattern noise associated with the detector for all targets, and the wavelength setting was adjusted according to the redshift of the target and was selected to cover major absorption line/emission line features including \cii, \civ, \siiie, and \siiv. The spectral resolutions of the observations vary among the three objects: objects J0906$+$56 and J0954$+$47 are point sources in the NUV acquisition images and their spectral resolution thus remains the nominal value of $\sim$15-20 \kms. The third object J1009$+$26, is extended in the NUV acquisition image and the spectral resolution is thus degraded. Following the method outlined in \citet{Henry2015}, we used the NUV acquisition image and the point source line spread function to derive the spectral resolution. Specifically, we first created a one-dimensional profile in the dispersion direction of the acquisition image, by summing the pixels in the cross dispersion direction. We then convolved this profile with the point source LSF to approximate the LSF for this spatially extended object and estimated its spectral resolution to be $\sim$30-40 \kms.

The raw data were processed and combined by the CALCOS pipeline v3.3.10. CALCOS corrects the data for instrumental effects, assigns a vacuum wavelength scale, and extracts flux-calibrated spectra. It applies a heliocentric correction to the final x1d files for each exposure, and combines the individual exposures to a single spectra.

\begin{figure}[!h]
    \centering
    \includegraphics[width=1\linewidth]{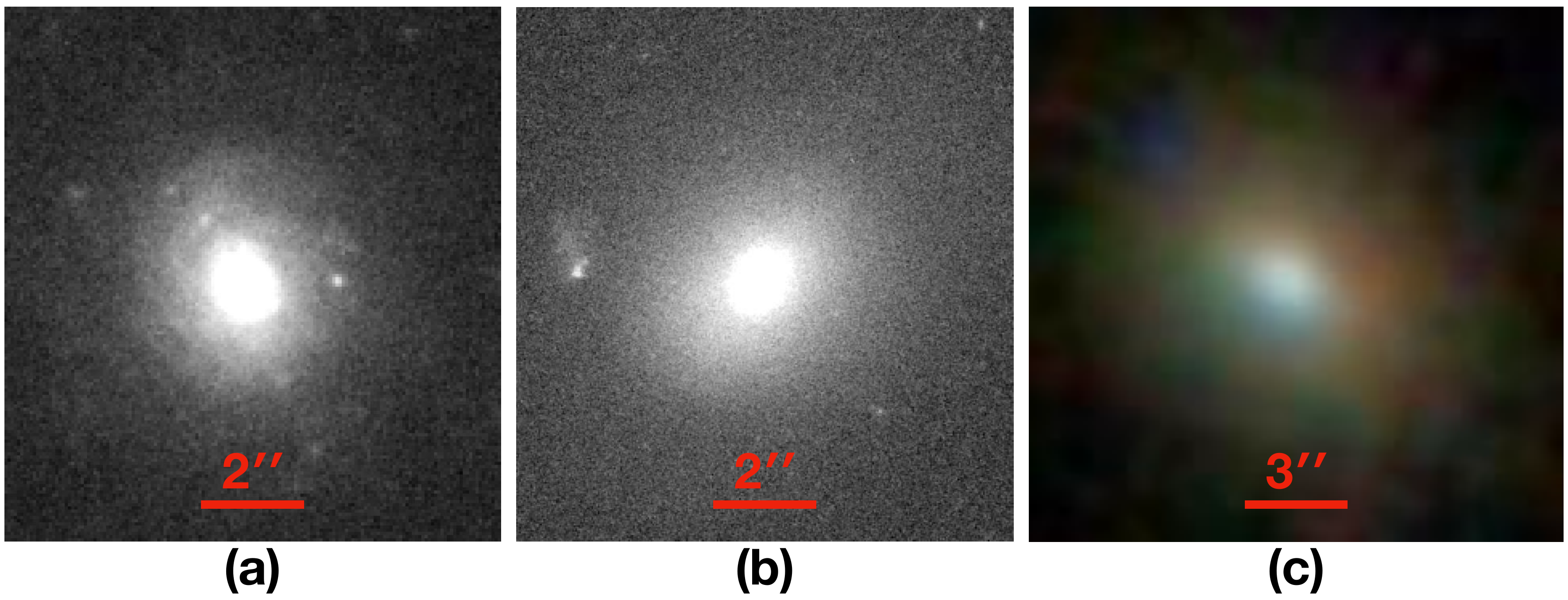}
    \caption{Archival \textit{HST} F606W images for sources J0906$+$56 (a) and J0954$+$47 (b), and SDSS composite image for J1009$+$26 (c). In each panel, the red horizontal bar indicates the corresponding spatial scale in arcseconds.}
    \label{fig:morphology}
\end{figure}

\subsection{Ancillary Data} \label{0522}

In addition to the \textit{HST}/COS data, we retrieve key physical properties of our sources from L20. Moreover, the circular velocity of the objects, \vcir\ are derived based on their stellar mass, adopting the baryonic Tully-Fisher relation calibrated in \citet{Reyes2011} as log$(v_{cir}) = 0.278 $log$(M_{*}) - 0.67$, with root-mean-square (RMS) residuals of 0.1 dex in \vcir. This relation was derived using spatially-resolved 2D gas kinematics in low-redshift emission-line galaxies where $v_{cir}  = \sqrt{2(v^{2}_{rot}+2\sigma^{2})}$. Here $v_{rot}$ and $\sigma$ are the rotation speed and mean value of the line-of-sight velocity dispersion, respectively.  These results are summarized in Table \ref{tab:targets}.

We also retrieve the archival HST F606W images (PID 13943; PI A. Reines) and SDSS images \citep{SDSSdr7} for our sources (Fig. \ref{fig:morphology}). The relevant HST data can be accessed via \dataset[10.17909/yp2h-yb46]{http://dx.doi.org/10.17909/yp2h-yb46}. The morphologies of our three sources are different. While J0906$+$56 is likely a late-type galaxy with possible spiral arms, J0954$+$47 and J1009$+$26 show no sign of spiral arms and are more consistent with early-type galaxies.

\begin{deluxetable*}{cccc cccccc}
\tablecolumns{10}
\tabletypesize{\scriptsize}
\tablecaption{Properties of the Objects\label{tab:targets}}
\tablehead{ \colhead{Name} & \colhead{Short Name} & \colhead{Redshift} & \colhead{log(M$_{\rm stellar}$)} &  \colhead{log(\vcir)} &
\colhead{R$_{50}$} &
\colhead{log(L$_{[OIII]}$)} & \colhead{log(L$_{AGN, [O III]}$)} & \colhead{log(L$_{AGN, FUV}$)} & \colhead{SFR}  \\
 \colhead{} & \colhead{} & \colhead{} & \colhead{[\msun]} &  \colhead{[\kms]} &
\colhead{[kpc]} &
\colhead{[erg s$^{-1}$]} & \colhead{[erg s$^{-1}$]} & \colhead{[erg s$^{-1}$]} & \colhead{[\msunyr]}  \\
\colhead{(1)} & \colhead{(2)} & \colhead{(3)} & \colhead{(4)} & \colhead{(5)} & \colhead{(6)} & \colhead{(7)} & \colhead{(8)} & \colhead{(9)} & \colhead{(10)} 
}
\startdata
SDSS J090613.75$+$561015.5 & J0906$+$56 & 0.0467 & 9.4 & 1.9$\pm{0.1}$ & 1.5 & 41.15$^{+0.01}_{-0.01}$  & 43.7 & 42.9 & $<$0.3 \\
SDSS J095418.16$+$471725.1& J0954$+$47 & 0.0327 & 9.1  & 1.9$\pm{0.1}$ & 2.0  & 41.36$^{+0.02}_{-0.02}$  & 43.9 & 42.8 & $<$0.3 \\
SDSS J100935.66$+$265648.9 & J1009$+$26 & 0.0145 & 8.8 & 1.8$\pm{0.1}$   & 0.7 & 40.48$^{+0.01}_{-0.01}$  & 43.0 & 42.0 & $<$0.1 \\
\enddata
\tablecomments{Column (1): SDSS name of the object; Column (2): Short name of the object used in this paper; Column (3): Redshift of the target measured from the stellar fit to the spectrum integrated over the KCWI data cube; Column (4): Stellar mass in unit of solar mass and in logarithm from the NASA Sloan Altas (\url{http://www.nsatlas.org/}); Column (5): Circular velocity in logarithm derived from stellar mass adopting calibration in \citet{Reyes2011}; Column (6): Half-light radius from the NSA, in unit of kpc;
Column (7): Total \oiii\ luminosity based on the observed total \oiii\ fluxes within the field of view of the KCWI data without extinction correction, in units of erg s$^{-1}$; Column (8): Bolometric AGN luminosity, based on the extinction-corrected [O~III] luminosity, in units of erg s$^{-1}$. Column (9): Bolometric AGN luminosity, based on the monochromatic continuum luminosity $L_{1320}$ from \textit{HST}/COS data and the CLOUDY AGN SED. See Appendix \ref{A1} for more details; Column (10): Upper limit on the star formation rate (SFR) based on the extinction-corrected \oiiab\ flux from the KCWI data, in units of \msunyr. Here we assume that 1$/$3 of the \oiiab\ emission is from the star formation activity, following \citet{Ho2005}.}
\end{deluxetable*}

\begin{figure}[!htb]   
\centering
\begin{minipage}[t]{0.4\textwidth}
 \centering
\includegraphics[width=\textwidth]{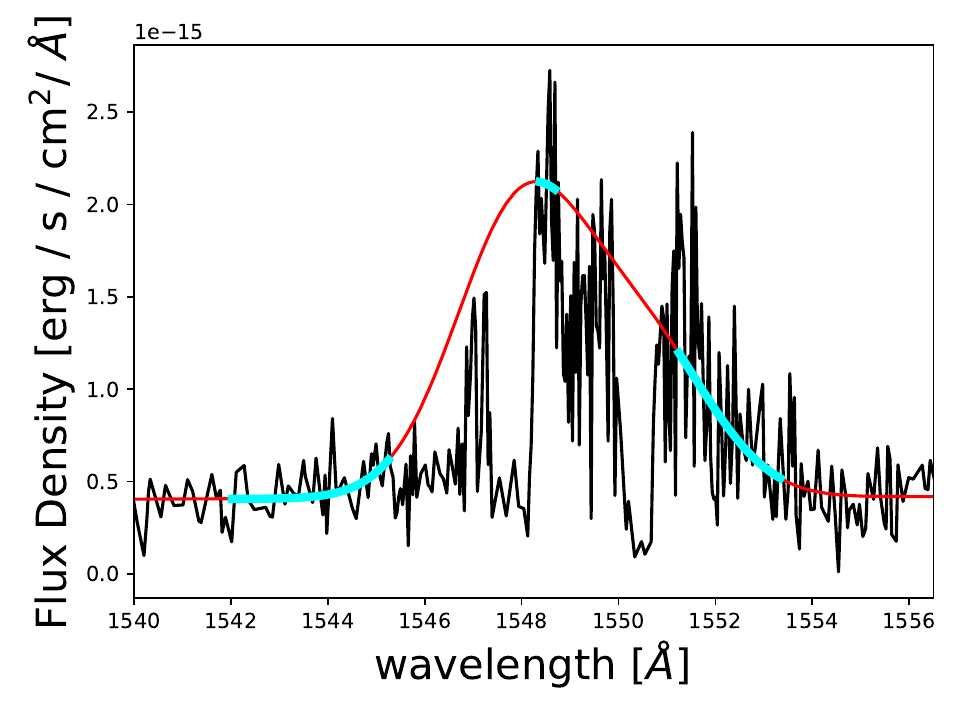}
\end{minipage}
 \begin{minipage}[t]{0.4\textwidth}
 \centering
\includegraphics[width=\textwidth]{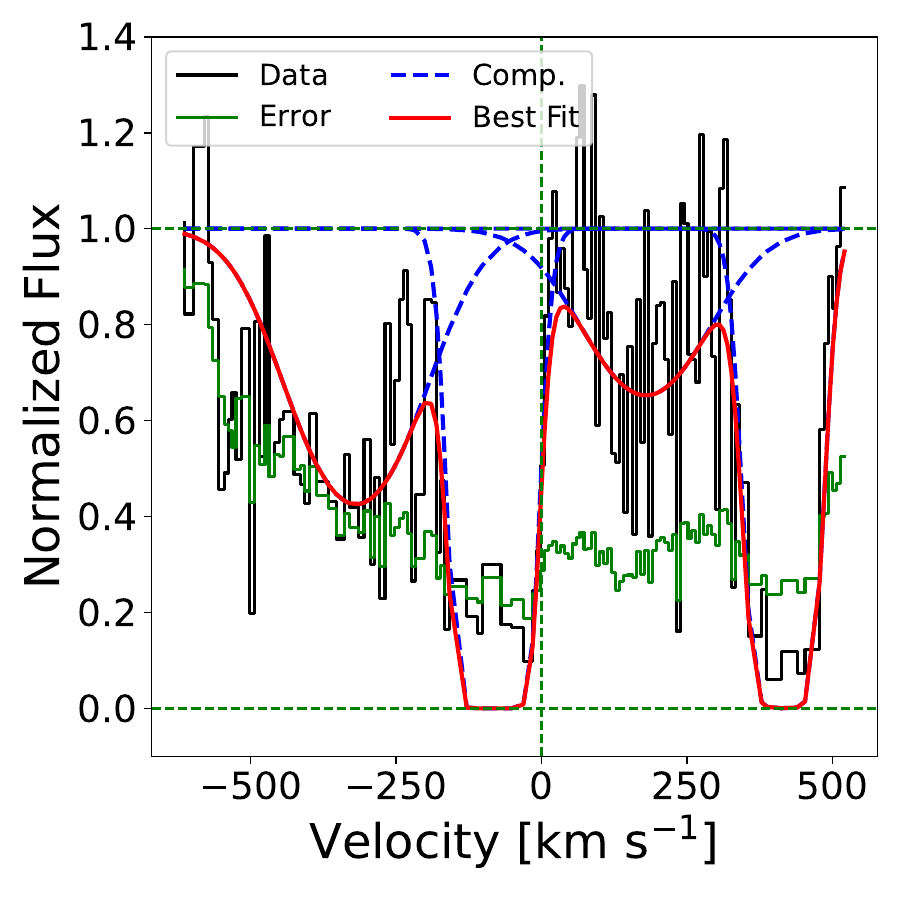}
\end{minipage}
\caption{The fit to the \civtext\ absorption + emission line complex in J0906$+$56. {Top:} The best-fit \civtext\ emission line doublet model (red) with two Gaussian components to recover the unabsorbed emission line profile. The fitting windows are marked on the best-fit model in cyan; {Bottom:} The best-fit model (red) of the absorption features normalized to the best-fit emission line model shown in the top panel. {The data is in black and the error is in green.} The individual Voigt profile components are shown as blue dashed lines. {The velocity is with respect to the systemic velocity of the bluer line} at the redshift listed in Table \ref{tab:targets}.}
\label{fig:J0906fit}
\end{figure}

\begin{figure*}[!htb]   
\centering
\begin{minipage}[t]{0.45\textwidth}
\includegraphics[width=\textwidth]{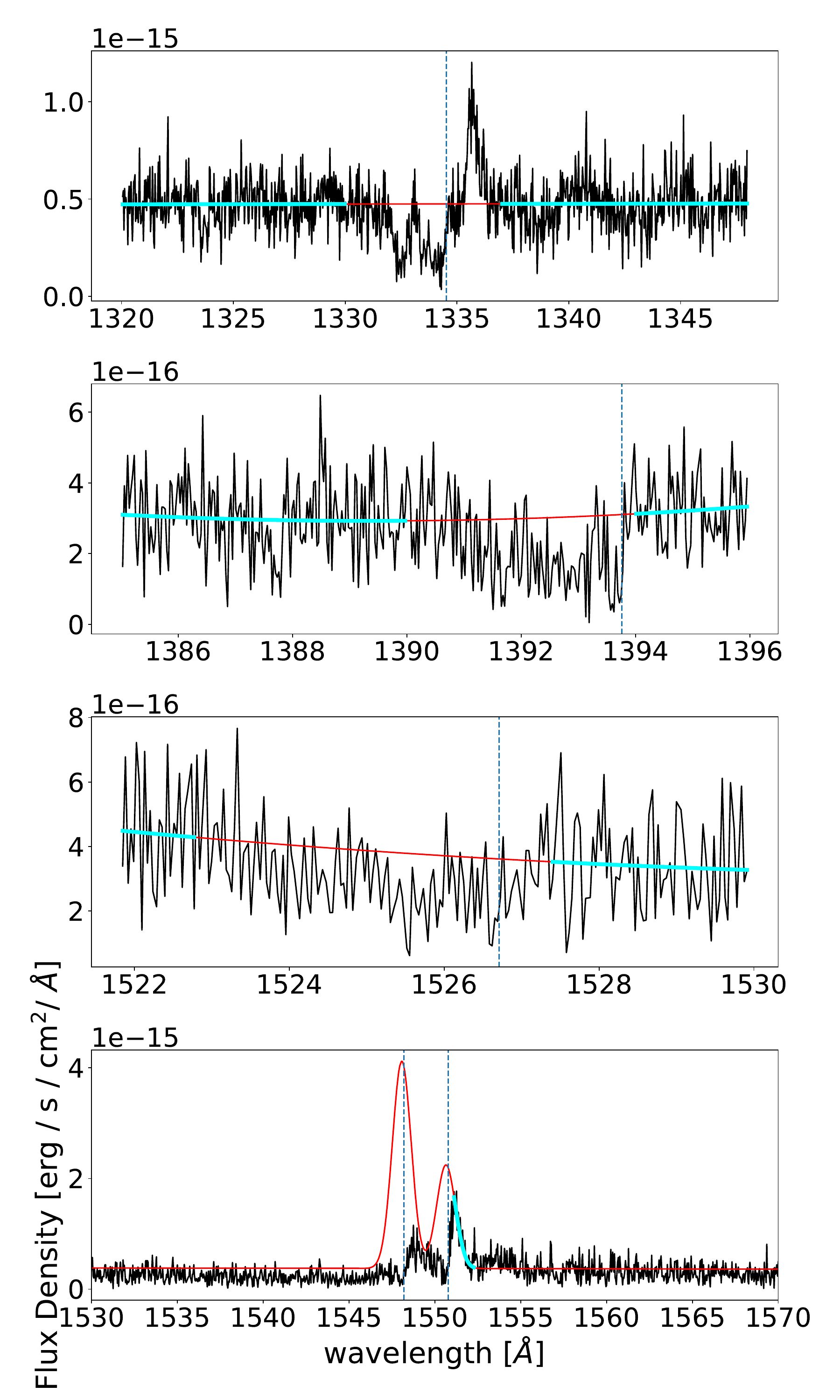}
\end{minipage}
\begin{minipage}[t]{0.45\textwidth}
\includegraphics[width=\textwidth]{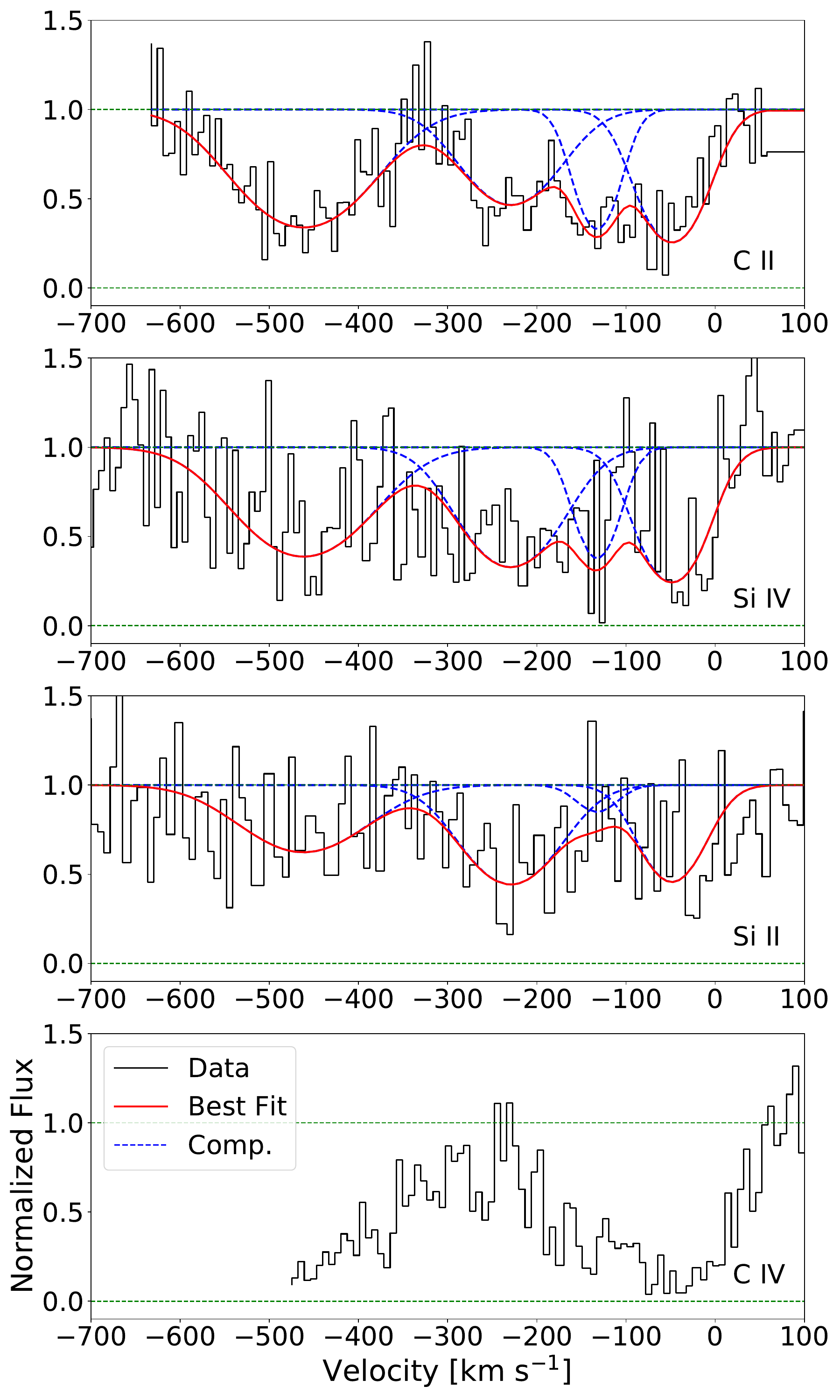}
\end{minipage}
\caption[The \cii, \siiva, \siiie, and \civ\ absorption features in J0954$+$47.]{The \cii, \siiva, \siiie, and \civb\ absorption features in J0954$+$47, from top to bottom. {Left:} The best-fit continuum/emission line model (red) in the vicinity of the absorption feature. The fitting windows are marked on the best-fit model in cyan. For \ciitext, the continuum is fit with a linear function. For \siivtext\ and \siiietext, the broad absorption trough in the vicinity of each absorption feature is fit with a 2nd-order polynomial. For \civtext, the intrinsic \civ\ emission line doublet is fit with two Gaussian profiles with the same centroid velocity and velocity dispersion, and a fixed flux ratio of 2. The centroid velocity and velocity dispersion are fixed to those of the
\heii\ emission in this object. The UV continuum is fit with a linear function to featureless regions beyond the plotting range of the figure. Note that the blueshifted, broad \civtext\ absorption trough and the possible, redshifted, broad \civtext\ emission line wing are not considered in our fit due to the large uncertainties. {Right:} The best-fit models (red) of the absorption features normalized by the best-fit continuum/emission line models shown in the left panel. The spectra are in the rest-frame of the corresponding absorption features. The individual Voigt profile components are shown as blue dashed lines. In the fit, the velocity and $b$ parameter (line width) of individual components for all three lines (\cii, \siiva, \siiie) are tied together. The \civb\ is excluded from the fit due to the large uncertainties in modeling intrinsic emission line$+$continuum profile, but the normalized absorption line profile is shown for visual comparison with the other three absorption features. The velocity is with respect to the systemic velocity at the redshift listed in Table \ref{tab:targets}.}
\label{fig:J0954fit4line}
\end{figure*}

\section{Outflows Traced by Blueshifted Absorption Features} \label{053}

\subsection {Detection and Characterization of Outflows} \label{0531}

{To characterize the absorption features in individual objects, we first fit the continua and unabsorbed emission line profiles in them with low-order polynomials and Gaussian functions convolved with the \textit{HST}/COS line-spread function (LSF)\footnote{Tabulated on the \textit{HST}/COS website \url{https://hst-docs.stsci.edu/cosihb}}, respectively. Next, the observed spectra are normalized with the best-fit continuum and intrinsic emission line models to obtain the normalized absorption line profiles. We then fit these normalized profiles with multiple velocity components with Voigt profiles, where the normalized intensity can be written as:}
{\begin{eqnarray}
I(v) = \prod\{1 - C_f [1-e^{-\tau(v)}]\} \\
\tau(v | N, b, z)  = N \sigma_0 f \Phi(v | b, z)
\label{eqn:absfit} 
\end{eqnarray}
$\tau$, $v$, $N$, $b$, $z$, $f$, and $\sigma_0$ are the optical depth, velocity, ion column density, Doppler parameter, redshift, oscillator strength, and cross section, respectively. $\Phi(v | b, z)$ is the normalized Voigt profile. The adopted atomic parameters are taken from \citet{Morton2003}. In all fits, the models are further convolved with the LSF, and the spectra are binned by 3 pixels for Nyquist sampling.
We adopted a customized software built on the non-linear least-squares fit implemented in \textit{LMFIT} \citep{lmfit} to search for the best-fit model for the absorption features. In our software, a velocity component (with Voigt profile) is added to the model if the Bayesian Information Criterion \citep[BIC;][]{bic} decreases, and this process stops when the minimum BIC value is found. The parameter uncertainties are calculated based on the output from the python package \textit{LMFIT}, which are the standard error from the estimated covariance matrix.
}

{None of the absorption features shows statistically significant evidence of partial covering (e.g. non-black saturated profiles). Specifically, while the bottoms of \civtext\ absorption troughs in J0906$+$56 (Fig. \ref{fig:J0906fit}) are above zero, they are consistent with the noise level of the observations. The best-fit model also gives a partial covering factor of 1 (i.e., no partial covering based on the current S/N of our spectrum). As a result, we fix the covering fraction to 1 in our fits to all objects and equation (1) is thus simplified to $I(v) = \prod[e^{-\tau(v)}]$.}

{Overall, objects J0906$+$56 and J0954$+$47 show clear evidence of outflows traced by absorption features. In object J1009$+$26, the detection is tentative as the blueshift of the absorption features is modest with large uncertainties. In the following, we describe the characteristic of the absorption features and fitting details for each object.}

{In J0906$+$56, only blueshifted \civ\ absorption features are detected on top of the broad, \civtext\ emission lines (Fig. \ref{fig:J0906fit} and Appendix \ref{A2}). We fit the featureless continuum with a linear function and the intrinsic \civtext\ emission line doublet with two Gaussian profiles, using fitting windows with no absorption signatures. The Gaussian profiles share the same centroid velocity and velocity dispersion, and their flux ratio is fixed to 2, the theoretical expectation in the optically thin limit. We also explore the case where the flux ratio of the emission line doublet is allowed to vary, but the final results remain similar. For the absorption feature, a total of two velocity components are required by the best-fit. The final best-fits for the continuum/emission line profile and the absorption line profiles are shown in Fig. \ref{fig:J0906fit}. }

\begin{figure}
    \centering
    \plotone{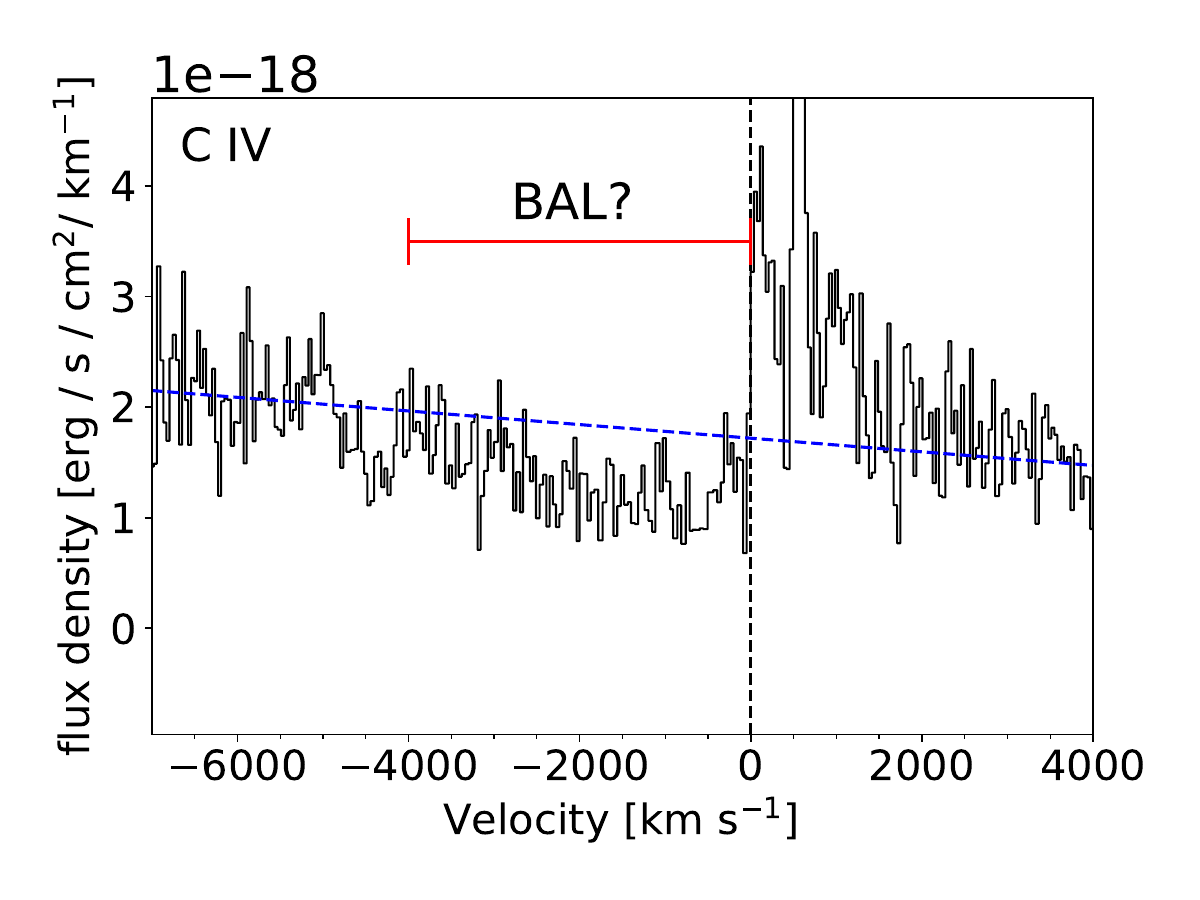}
    \caption{Zoom-in view on the potential \civtext\ BAL feature in J0954$+$47. The zero velocity of \civa\ corresponds to the redshift listed in Table \ref{tab:targets} and is denoted by the vertical black line. The best-fit continuum with a linear function is indicated by the blue dashed line. The red horizontal bar denotes the breadth of the potential BAL feature.}
    \label{fig:BAL}
\end{figure}

{In J0954$+$47, blueshifted absorption features with similar profiles are seen in \cii, \civ, \siiie, \siiv\ and possibly \oiuv\ (Fig. \ref{fig:J0954fit4line} and Appendix \ref{A2}). 
Additionally, much broader ($\sim$4000 \kms) absorption troughs are seen for both \civtext\ and \siivtext\ doublets (but not for other lower ionization transitions; e.g., see Fig. \ref{fig:BAL} for \civtext\ as an example). The line widths of these broad troughs fall in the typical range of those of broad absorption lines (BAL) seen in luminous quasars \citep[e.g.][]{Weymann1991}. If these broad troughs are indeed BAL features as in luminous quasars, it would be the first BAL ever discovered in a dwarf AGN to our knowledge, with a balnicity \citep{Weymann1991} of $\sim$740 \kms. Nevertheless, these absorption troughs are possibly accompanied by redshifted broad emission line features reaching $\sim$2000 \kms\ (with peak S/N $\sim$ 3). The entire features may thus also be P-Cygni-like profiles and instead trace a spatially unresolved stellar wind in this object.}

{To further characterize the narrow absorption features (i.e., excluding the potential BALs), we first focus on the \cii, \siiva, and \siiie\ lines. These absorption features are either not or only mildly contaminated by nearby emission and/or broad absorption features (Appendix \ref{A2}) and thus the derived normalized absorption line profiles are relatively more robust. Specifically, the continuum near \cii\ is fit with a linear function. The broad absorption trough in the vicinity of \siiva\ and \siiie\ are both fit with 2nd-order polynomials. Next, for the \civb\ absorption feature, we aim to obtain an approximate profile of it as it is blended with the nearby \civtext\ doublet emission, \civa\ absorption, and broad \civtext\ BAL-like trough. The intrinsic \civ\ emission line is fit with two Gaussian profiles with the same centroid velocity and velocity dispersion, and a fixed flux ratio of 2. The fitting window is chosen as the presumably pure \civtext\ emission-line wing, and the centroid velocity and velocity dispersion are fixed to those of the \heii\ emission line. As we only care about the spectral range of $\gtrsim-$500 \kms\ of \civb\ where the contribution from the broad \civtext\ BAL-like trough should be trivial, this broad trough is thus not modeled in the fit.  Finally, for the \civa\ and \siivb\ absorption features, the nearby emission/absorption features are more complex than those near \civb\ and our attempt to constrain the intrinsic absorption line profiles results in large uncertainties. These two features are thus not considered in the following analysis.}

{We then obtain normalized absorption line profiles for \ciitext, \siivtext\ and \siiietext\ based on the best-fit continuum/emission line profiles above, and fit them simultaneously where the centroid velocities and velocity dispersions (i.e., Doppler parameters) of the same velocity components are tied together. A total of four velocity components are required by the best-fit. The final best-fits for the continuum/emission line profiles and the absorption line profiles are shown in Fig. \ref{fig:J0954fit4line}.}

{In J1009$+$26, the overall featureless continuum is fit with a linear function. The unabsorbed \civtext\ emission line profile is modeled in the same way as that for J0954$+$47, except that the line width of \civtext\ is not fixed to that of \heiitext\ but set as a free parameter instead. The recovered \civtext\ absorption features are relatively weak and poorly constrained due to the large uncertainties of the unabsorbed \civtext\ emission line profile.}

\begin{figure*}[!htb]   
\centering
\begin{minipage}[t]{0.45\textwidth}
\includegraphics[width=\textwidth]{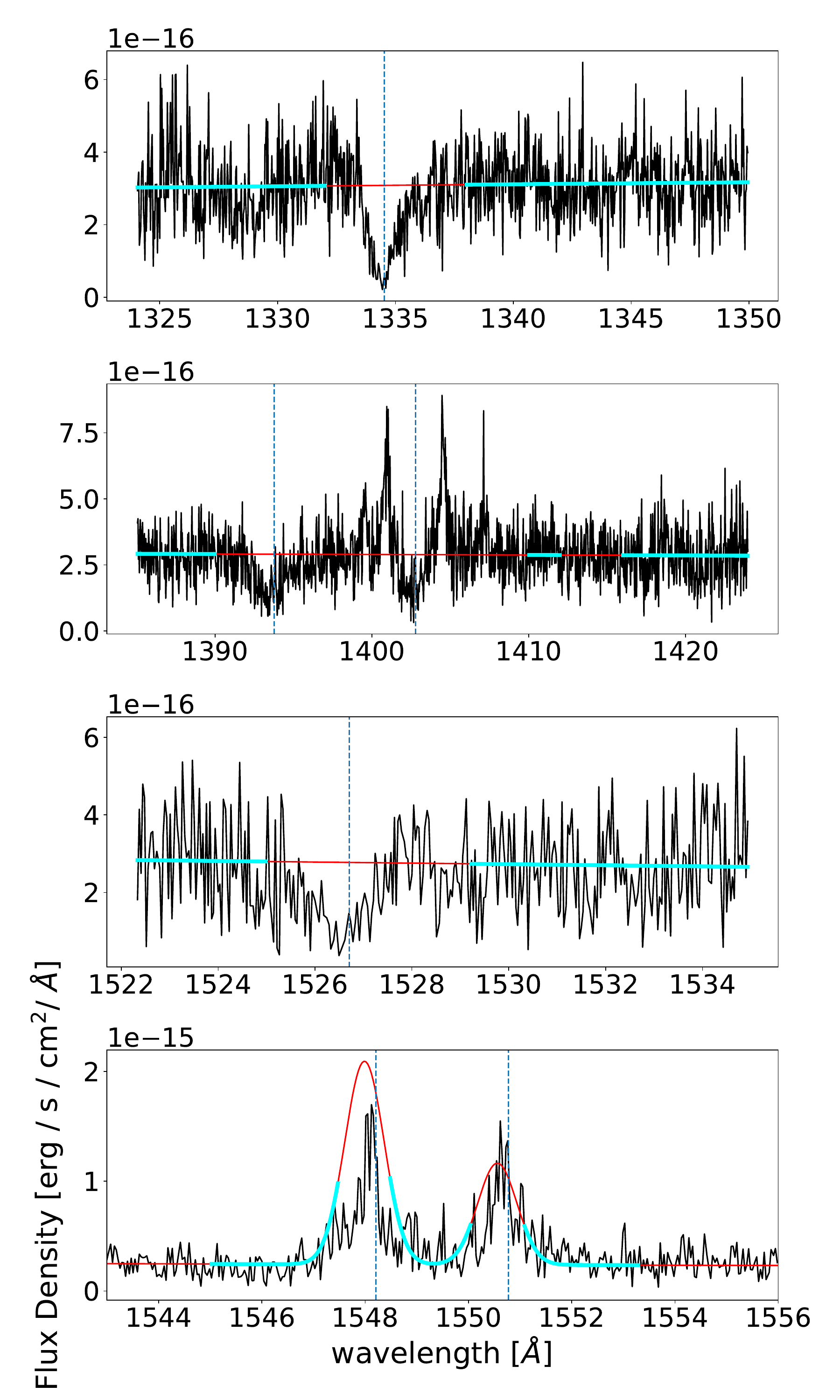}
\end{minipage}
\begin{minipage}[t]{0.45\textwidth}
\includegraphics[width=\textwidth]{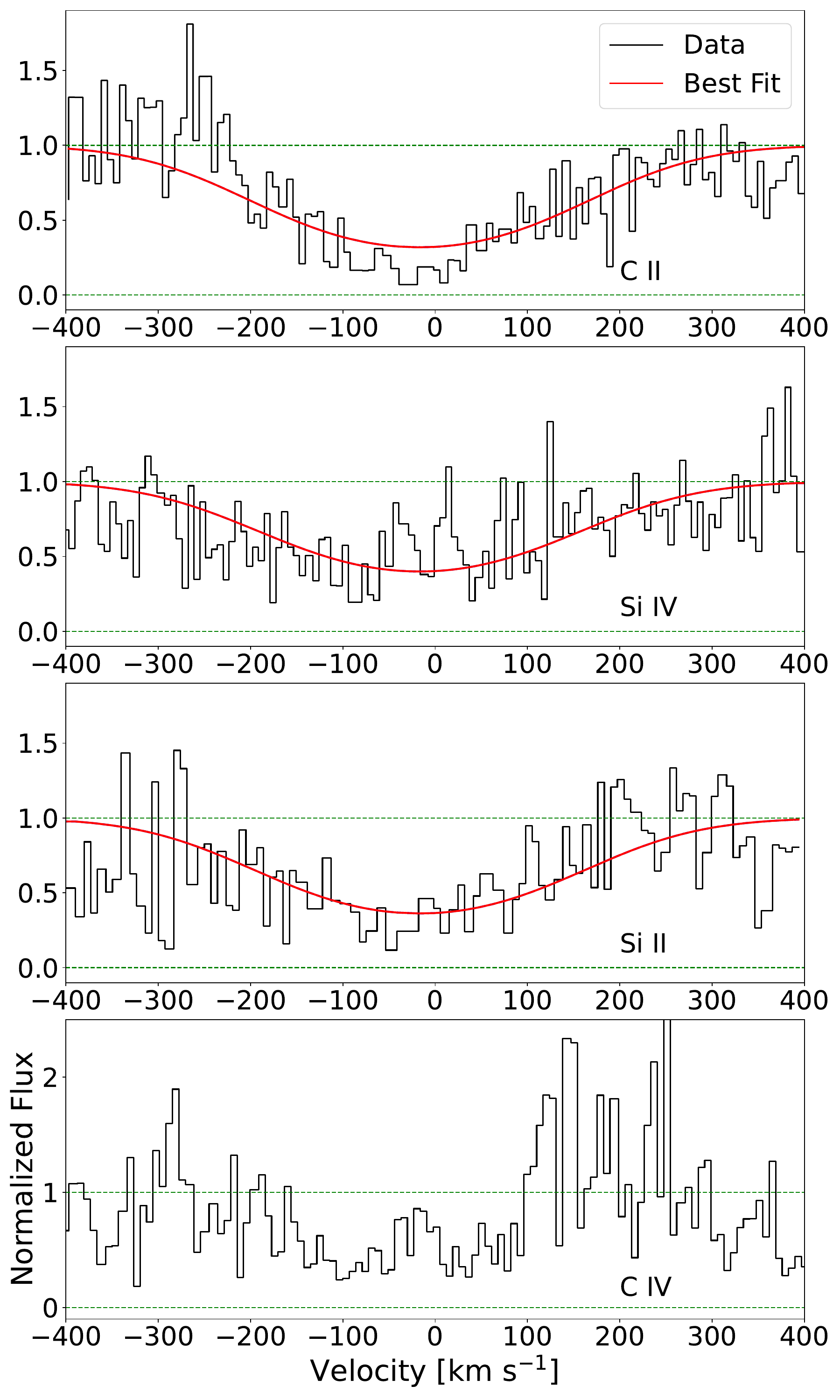}
\end{minipage}
\caption[]{The \cii, \siiv, \siiie, and \civ\ absorption features in J1009$+$26, from top to bottom. {Left:} The best-fit continuum/emission line model (red) in the vicinity of the absorption feature. The fitting windows are marked on the best-fit model in cyan. The featureless continuum is fit with a linear function to selected wavelength windows marked in cyan. The intrinsic \civ\ emission line doublet is fit with two Gaussian profiles with the same centroid velocity and velocity dispersion, and a fixed flux ratio of 2. The location of the corresponding transitions at the systemic velocity are indicated by blue vertical dashed lines. {Right:} The best-fit models (red) of the absorption features normalized by the best-fit continuum/emission line models shown in the left panel. The \civtext\ feature is excluded from the fit due to the large uncertainties in modeling the unabsorbed emission line profile, but the normalized absorption line profile is shown for visual comparison with the other three absorption features. In the fit, the velocity and $b$ parameter (line width) of individual components for all lines are tied together. For \siivtext\ and \civtext, only the bluer line of the doublet is shown for a zoom-in view of the absorption line profile. The velocity is with respect to the systemic velocity at the redshift listed in Table \ref{tab:targets}.}
\label{fig:J1009fit4line}
\end{figure*}

{We then examine the normalized absorption features closely and find that the blueshift of absorption features is in general modest (Fig. \ref{fig:J1009fit4line} and Appendix \ref{A2}). \siiv\ and \siiie\ absorption features are mildly blueshifted with \vwu\ $\sim-$20 \kms\ when each feature is fit individually with a single Voigt profile (Here \vwu\ is the velocity at the location that accumulates 50\% of the total equivalent width (EW) of the entire absorption feature). Blueshifted \civ\ absorption features at similar velocities may also be present, despite the large uncertainties resulting from the modeling of unabsorbed \civtext\ emission line. 
However, the \cii\ absorption feature is not blueshifted when fit alone with a Voigt profile and is detected only at the systemic velocity with a relatively large uncertainty (\vwu\ $\sim 0\pm{30}$ \kms). Moreover, when \siiitext, \siivtext\ and \ciitext\ absorption features are fit together with the same centroid velocity and Doppler parameter, the blueshift is small with best-fit \vwu\ $\sim -15\pm{15}$ \kms. Additionally, the absorption line profiles are smooth and broad (Doppler parameter b$\sim$190 \kms), which are different from the profiles typically observed in intervening and ISM absorbers in star-forming galaxies \citep[e.g.,][]{Werk2013,Savage1996} but similar to intrinsic AGN absorbers often associated with outflows \citep[e.g.,][]{Hamann1999}. 
Overall, we cannot completely rule out the possibility that the absorption features are tracing the kinematically quiescent gas at/near systemic velocity and thus the outflow in this object is only marginally detected.}

{In our final fit to the normalized absorption features, we adopt the simultaneously fit to \ciitext, \siivtext\ and \siiietext\ where the centroid velocities and Doppler parameters of the same velocity components are tied together. We find that only one velocity component is required by the best-fit, and the final best-fits for the continuum/emission line profiles and the absorption line profiles are shown in Fig. \ref{fig:J1009fit4line}.}

{The final results from the best-fits of absorption features for all three objects are summarized in Table \ref{tab:results}.}

\subsection {The Galactic-scale Outflow in J0954$+$47} \label{0532}

The blueshifted absorption features of various ions in J0954$+$47 share similar velocity distributions (see Fig. \ref{fig:J0954fit4line}), tracing the same fast outflow within this object (The potential \civtext\ and \siivtext\ BALs shown in Fig. \ref{fig:BAL} and described in Section \ref{0531} are not considered here due to the lack of constraints on its radial distance based on our data). The lower limit on the ratio of \cii\ and \ciiex\ lines (\ciiex\ undetected) also provides tight constraint on the electron density and thus the radial distance of the outflow. In the remainder of this section, we take advantage of these constraints to derive the physical properties of the outflow in J0954$+$47 and evaluate the power source and impact of the outflow. Note that we refer to the entire absorption feature as the ``entire outflow'', and the most blueshifted velocity component in the best-fit model shown in Fig. \ref{fig:J0954fit4line} as the ``high-v outflow'' hereafter. Such analysis is not reliable and thus not carried out for J0906$+$56 and J1009$+$26, since the outflow is only detected in \civtext\ for the former and is only tentatively detected for the latter.

\begin{deluxetable*}{cccccc ccccc}
\tablecolumns{11}
\tablecaption{Properties of the Outflows\label{tab:outflow}}
\tabletypesize{\scriptsize}
\tablehead{ 
\colhead{Object} & \colhead{Absorber} & \colhead{Comp.} & \colhead{\vwu} & \colhead{{$b$}} & \colhead{$log(N_{ion}$)} & \colhead{$log(N_{H}$)} & \colhead{$R$} &
\colhead{$log(\dot{m}_{out})$} &
\colhead{$log(\dot{p}_{out})$} & \colhead{$log(\dot{E}_{out})$}   \\
 &  &  & \colhead{[\kms]} & \colhead{[{\kms}]}& \colhead{[$cm^{-2}$]} & \colhead{[$cm^{-2}$]} & \colhead{[kpc]} &
\colhead{[\msunyr]} &
\colhead{[$dynes$]} & \colhead{[$erg\ s^{-1}$]} \\
\colhead{(1)} & \colhead{(2)} & \colhead{(3)} & \colhead{(4)} & \colhead{{(5)}} & \colhead{(6)} & \colhead{(7)} & \colhead{(8)} & \colhead{(9)} & \colhead{(10)} & \colhead{(11)}
}
\startdata
\multirow{2}{*}{J0906$+$56} & \multirow{2}{*}{\civtext} & entire & $-130\pm{40}$ & $180\pm{20}$, $60\pm{10}$  & $15.3\pm{0.1}$ & ... & ... & ... & ... & ...   \\
                            &                           & high-v & $-320\pm{30}$ & $180\pm{20}$                         & $14.4\pm{0.1}$ & ... & ... & ... & ... & ...   \\
\hline
\multirow{6}{*}{J0954$+$47} &  \ciitext   & \multirow{3}{*}{entire}  & \multirow{3}{*}{$-260\pm{20}$}  &  \multirow{3}{*}{$100\pm{10}$, $60\pm{12}$, $85\pm{30}$, $50\pm{10}$} & $15.0\pm{0.1}$  & \multirow{3}{*}{$19.7\pm{0.1}$} & \multirow{3}{*}{$0.49\pm{0.04}$} & \multirow{3}{*}{$-0.1\pm{0.2}$} & \multirow{3}{*}{$33.2\pm{0.2}$} & \multirow{3}{*}{$40.6\pm{0.2}$} \\
                            &  \siivtext  &                          &                                 &                                                                    &  $14.4\pm{0.1}$ &                                 &                                  &                                 &                                 &  \\
                            &  \siiietext &                          &                                 &                                                                    &  $14.7\pm{0.2}$ &                                 &                                  &                                 &                                 &  \\
                            &  \ciitext   & \multirow{3}{*}{high-v}  & \multirow{3}{*}{$-460\pm{10}$}  &  \multirow{3}{*}{$100\pm{10}$}                                     & $14.60\pm{0.04}$ & \multirow{3}{*}{$19.20\pm{0.06}$} & \multirow{3}{*}{$0.52\pm{0.03}$} & \multirow{3}{*}{$-0.3\pm{0.3}$} & \multirow{3}{*}{$33.2\pm{0.3}$} & \multirow{3}{*}{$40.9\pm{0.3}$} \\
&   \siivtext  &         &                      &                          & $13.90\pm{0.07}$ &  &  &  &  &  \\
&  \siiietext &           &                    &                          & $14.10\pm{0.15}$ &  &  &  &  & \\ 
\hline
\multirow{3}{*}{J1009$+$26} &  \ciitext   & \multirow{3}{*}{entire} & \multirow{3}{*}{$-15\pm{15}$}  & \multirow{3}{*}{$190\pm{10}$} & $14.9\pm{0.1}$ & \multirow{3}{*}{...} & \multirow{3}{*}{...} & \multirow{3}{*}{...} & \multirow{3}{*}{...} & \multirow{3}{*}{...} \\
 &  \siivtext   & & & \  & $14.2\pm{0.1}$ &  &  &  &  &  \\
&   \siiietext  &  &                             &                          & $14.8\pm{0.1}$ &  &  &  &  &  \\
\enddata
\tablecomments{Column (1): Object name; Column (2): Ion considered; Column (3): Velocity component considered in the calculation. The ``entire'' label indicates the values for the entire absorption feature. Specifically, we adopt this \vwu\ as the outflow velocity for the calculation of mass, momentum and kinetic energy outflow rate of the entire outflow in column (9)--(11). So these are effectively average outflow rates. The ``high-v'' label indicates the component with the highest outflow velocity in the best-fit model. {Column (4): Velocity at the location that accumulates 50\% of the total equivalent width (EW) of the entire absorption feature; Column (5): Doppler parameter from the best-fits. For the ``entire component'', we list the value for each individual component separately;} Column (6): Ion column density in logarithm; Column (7): Hydrogen column density in logarithm derived from Section \ref{05321}; Column (8): Outflow radial distance in kpc derived from Section \ref{05321}; Column (9): Mass outflow rate in logarithm; Column (10): Momentum outflow rate  in logarithm; Column (11): Kinetic energy outflow rate in logarithm.}
\label{tab:results}
\end{deluxetable*}

\subsubsection{Photoionization Modeling} \label{05321}

\begin{figure*}[!htb]
\begin{minipage}[t]{0.5\textwidth}
\includegraphics[width=\textwidth]{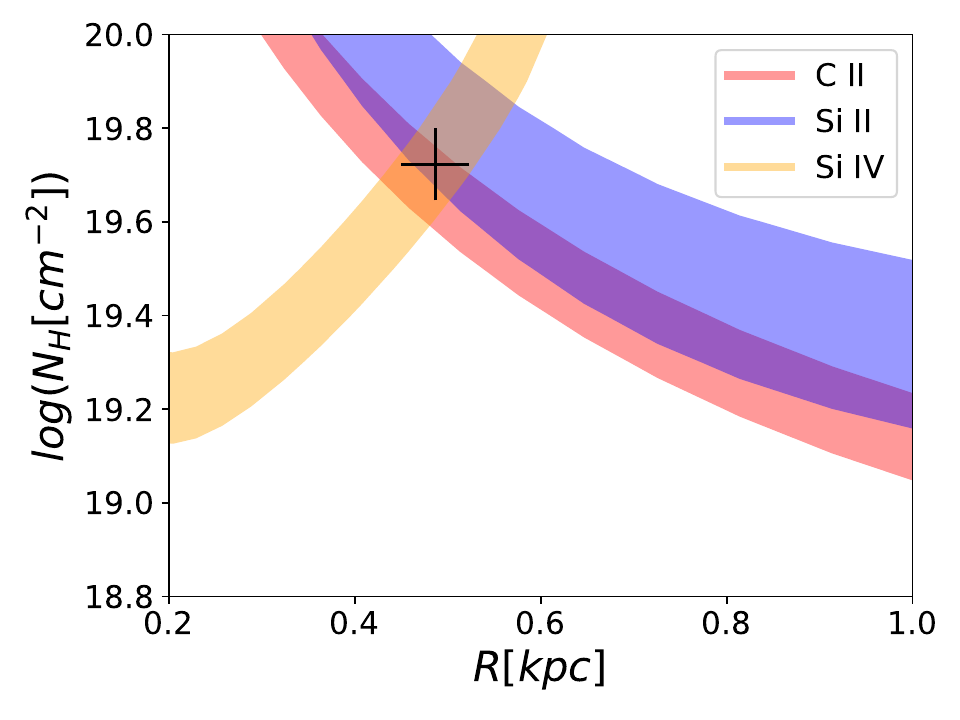}
\end{minipage}
 \begin{minipage}[t]{0.5\textwidth}
\includegraphics[width=\textwidth]{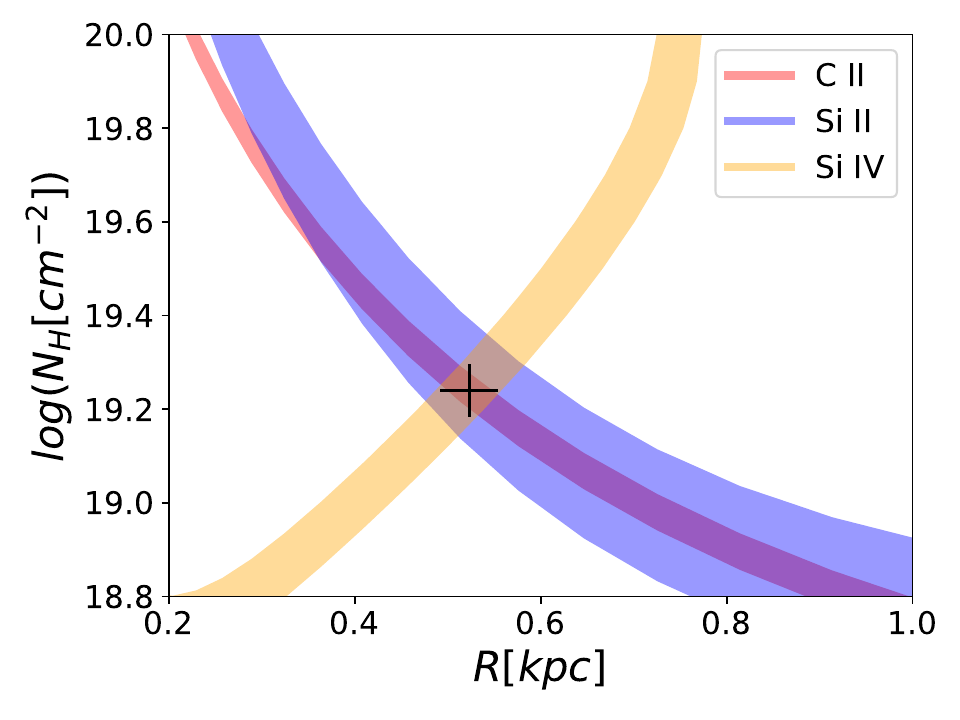}
\end{minipage}
\caption[Phase plot showing the ionization solutions for the entire outflow and the high-v outflow of J0954$+$47.]{Phase plot showing the ionization solutions for the entire outflow ({left}) and the high-v outflow ({right}) of J0954$+$47. Each colored contour represents the locus of models (outflow radial distance $R$, hydrogen column density $N_H$) which predicts a ion column density consistent with the observed value of that ion. The width of each locus corresponds to the 1-$\sigma$ uncertainty in the observed ion column density. The center of the black cross indicates the best solution based on the \chitwo\ minimization and the size of the cross indicated the corresponding uncertainties. See Section \ref{05321} for more details.}
\label{fig:J0954phase}
\end{figure*}

We adopt CLOUDY \citep[C22.00][]{CLOUDY} to model the properties of both the entire outflow and the high-v outflow adopting the measured column density of the outflow components of \ciitext, \siiietext, and \siiva, assuming that they trace the same outflowing material. The \oiuv\ absorption feature is too weak (S/N$\sim$2.5) to be measured robustly, and we do not include it in our modeling described below.

We set up the modeling as follows: we assume that the ionizing spectrum is a typical AGN continuum described in Section 6.2 of the CLOUDY manual\footnote{\url{https://gitlab.nublado.org/cloudy/cloudy/-/wikis/home}}, which gives $f_\nu = \nu^{\alpha_{uv}} exp(-h\nu/kT_{BB})exp(-kT_{IR}/hv)+a\nu^{\alpha_x}$. $T_{BB}$ is the temperature of the ``Big Blue Bump'' component of the AGN, a rising power law with a high-energy exponential cut-off. $\alpha_{uv}$ is the low-energy slope of the ``Big Bump'' component. $\alpha_x$ is the slope of the X-ray component. The coefficient $a$ is adjusted to produce the correct \aox. We normalize this AGN continuum at 1320 \AA\ to the observed flux density of the COS spectrum. We also set the X-ray to UV ratio, \aox, to $-$1.9 as measured from \citet{dwarfXrayUV}. We set the other three free parameters, $T_{BB}$, $\alpha_{uv}$, and $\alpha_x$ provided by CLOUDY to their default values (1.5$\times$10$^{5}$ K, $-$0.5 and $-$1, respectively, which are typical for AGN). Two more parameters are needed for the modeling, the electron density \nel\ and metal abundance. Due to the absence of \ciiex\ absorption, we obtain an upper limit of $\sim$ 10 cm$^{-3}$ for \nel\ adopting the upper limit on the \ciiext/\ciitext\ ratio\footnote{The required atomic data are retrieved from CHIANTI atomic database for astronomical spectroscopy via the Python interface ChiantiPy}. On the other hand, we do not expect \nel\ to be much less than 10 cm$^{-3}$, given that the typical \nel\ for outflows in star-forming dwarf galaxies is $\gtrsim$ 10 cm$^{-3}$ \citep[e.g.][]{Xu2022}. Therefore, we set {\nel\ to 10 cm$^{-3}$} in the following analysis. As for the metal abundance, we set it to solar, the same as the values obtained from the photoionization modeling of the optical emission lines (L20).

With the model set-up above, the column density of each absorption feature (with its 1-$\sigma$ uncertainty) defines a locus in the 2D diagram of total hydrogen column density \NH\ versus outflow radial distance $R$\footnote{The default independent variable in CLOUDY is the ionization parameter $U=\frac{Q}{4\pi c n_e R^2}$. With a given $n_e$, we convert $U$ into R.}. Following the same approach adopted in previous work \citep[e.g.][]{Borguet2012a,Arav2013}, the best-fit total hydrogen column density \NH\ and the outflow radial distance $R$ is then obtained by minimizing the \chitwo:

\begin{eqnarray}
\chi^2 = \sum_{i}(\frac{logN_{i,mod} - logN_{o,obs}}{logN_{i,obs}-log(N_{i,obs}\pm{\sigma_i})})^2
\end{eqnarray}
where, for ion $i$, $N_{i,obs}$ and $N_{i,mod}$ are the observed and modeled column densities, and $\sigma_{i}$ is the error in the measured column density. 

The best-fit \NH\ and $R$ are $19.7\pm{0.1}$ cm$^{-2}$ and $0.49\pm{0.04}$ kpc for the entire outflow and $19.20\pm{0.06}$ cm$^{-2}$ and $0.52\pm{0.03}$ kpc for the high-v outflow, respectively. The 1-$\sigma$ errors are estimated as the full x and y extent of the overlapping region of the three loci in Fig.\ \ref{fig:J0954phase}. These results are listed in Table \ref{tab:outflow}.

\subsubsection{Energetics} \label{05322}

With the \NH\ and $R$ obtained above, the mass, momentum and kinetic energy outflow rate can then be calculated, using the equations:

\begin{eqnarray}
\dot{m}_{out} = \Omega N_{H} \mu m_p R v \\
\dot{p}_{out} = \Omega N_{H} \mu m_p R v^2 \\
\dot{E}_{out} = \frac{1}{2}\Omega N_{H} \mu m_p R v^3
\label{eqn:KE2} 
\end{eqnarray}

In the equations above, the outflow velocity $v$ is set to \vwu\ (velocity at half the equivalent width) of the best-fit absorption line profile. \footnote{{Note that we are effectively calculating an average for the entire outflow by adopting this definition of outflow velocity, and assuming that terms $v(r)n(r)r$, $[v(r)]^2n(r)r$ and $[v(r)]^3n(r)r$ are constant in the integral $\int_{0}^{4\pi}\int_{0}^{R}v(r)n(r)r drd\Omega$, $\int_{0}^{4\pi}\int_{0}^{R}[v(r)]^2n(r)r drd\Omega$, and $\int_{0}^{4\pi}\int_{0}^{R}[v(r)]^3n(r)r drd\Omega$ to obtain $\dot{m}_{out}$, $\dot{p}_{out}$, and $\dot{E}_{out}$, respectively. This leads to the $N_{H} R v$, $N_{H} R v^2$, and $N_{H} R v^3$ terms in equations (4)--(6)}.}.  $\Omega$ is the solid angle subtended by the outflow as seen from the center. While $\Omega$ cannot be measured directly from our data, the outflow likely has a wide opening angle: the outflow seen in the optical emission lines, which should be physically connected with the outflow seen in absorption lines, favors such geometry (L20). In addition, no evidence of partial covering is present in the absorption line profiles. Therefore, we set $\Omega = 4\pi$ in our calculations. The final results are recorded in Table \ref{tab:outflow}. Note again that as stated in Section \ref{05321}, it is possible that the actual \nel\ of the outflow is smaller than the 10 cm$^{-3}$ adopted, so the  outflow radial distance and outflow rates could be even larger than listed in Table \ref{tab:outflow}.

The spatial extent and energetics of the outflow derived from the absorption features are consistent with those measured from the blueshifted \oiii\ emission line (L20; $R < 1.6$ kpc and log~$\dot{E}_{out} > 39.6$ erg s$^{-1}$). This suggests that the blueshifted absorption and emission features are tracing the same galactic outflow in this object, as assumed in the above calculations.

\begin{figure*}
\centering
\begin{minipage}[t]{0.48\textwidth}
\includegraphics[width=\textwidth]{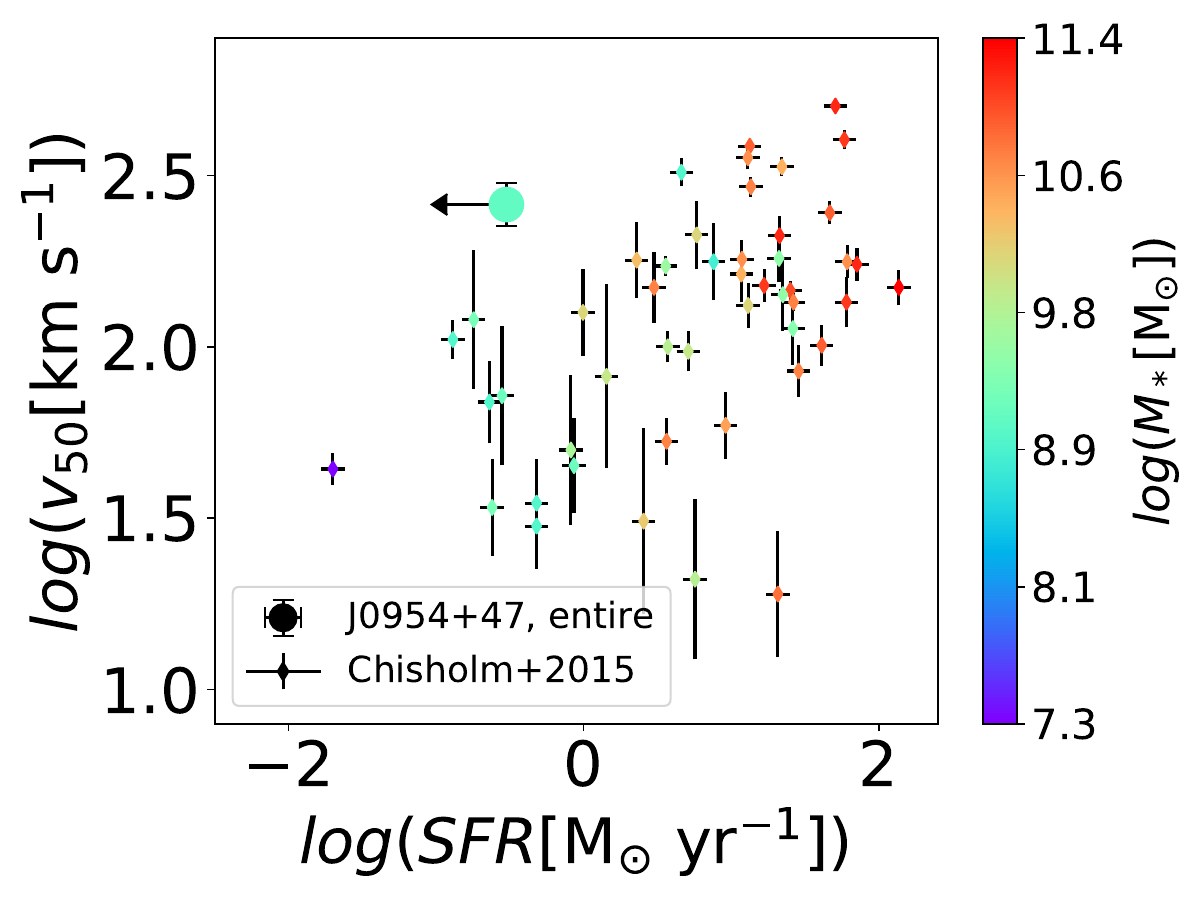}
\end{minipage}
\begin{minipage}[t]{0.48\textwidth}
\includegraphics[width=\textwidth]{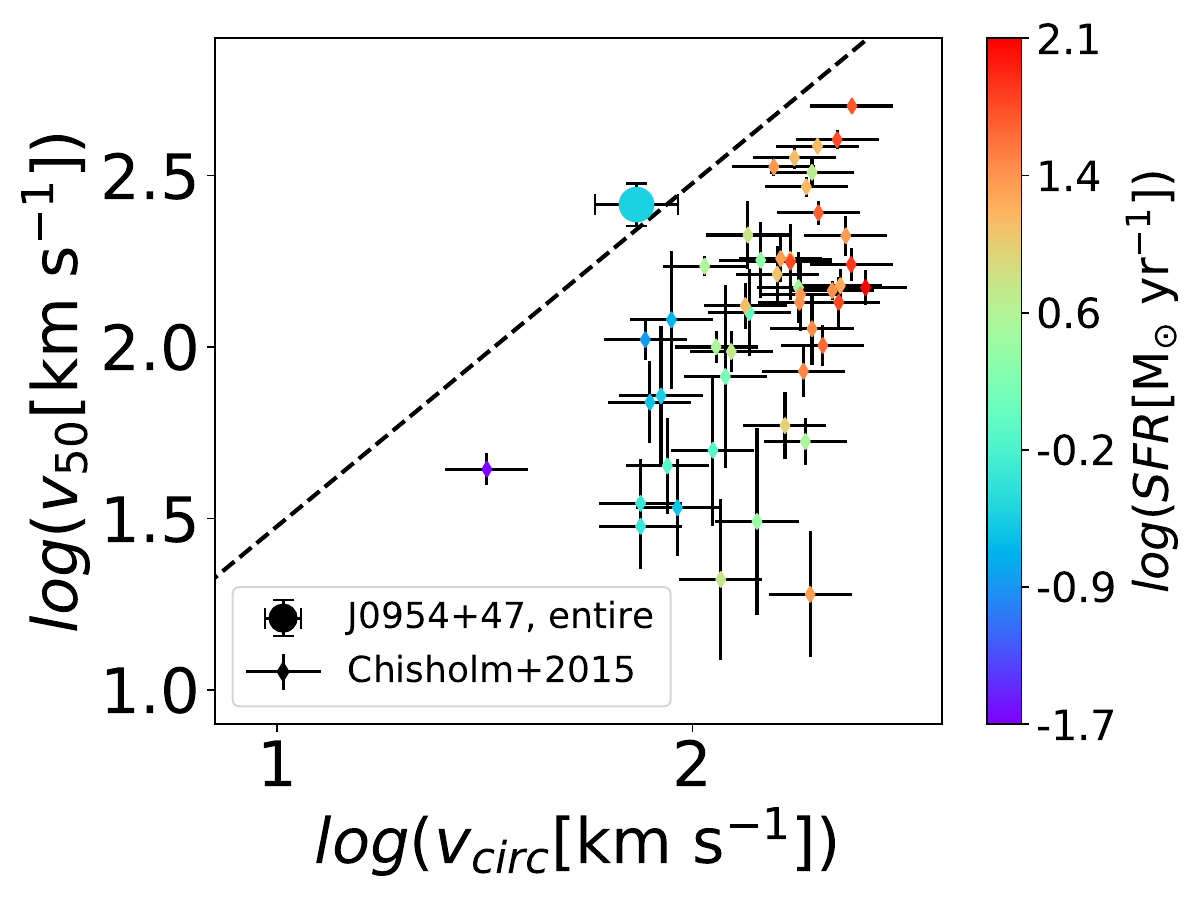}
\end{minipage}

\begin{minipage}[t]{0.48\textwidth}
\includegraphics[width=\textwidth]{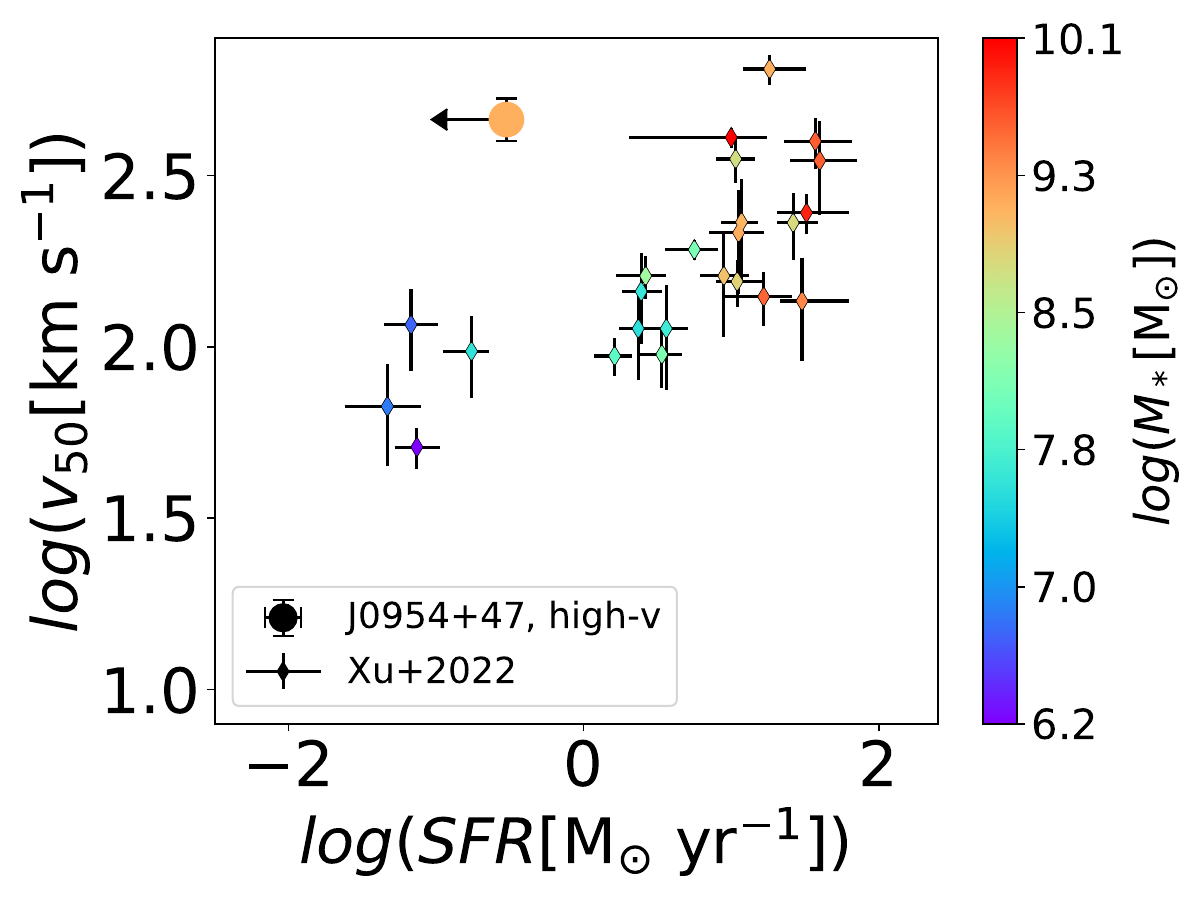}
\end{minipage}
\begin{minipage}[t]{0.48\textwidth}
\includegraphics[width=\textwidth]{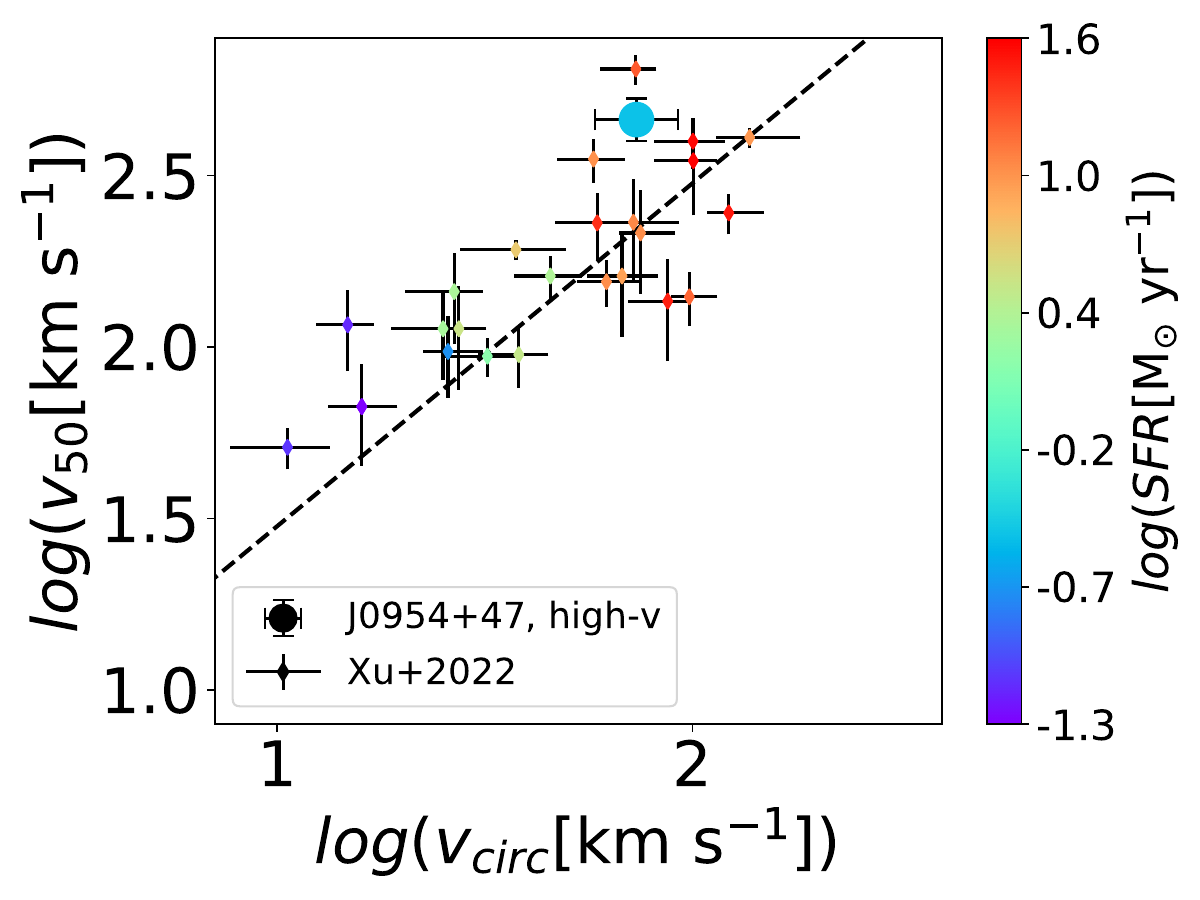}
\end{minipage}
\caption[Outflow velocity (\vwu) as a function of SFR (left) and circular velocity (right).]{Outflow velocity (\vwu) as a function of SFR (left panels) and circular velocity (right panels) based on the overall absorption feature (entire outflow; upper panels) and the highest velocity component in the best-fit (high-v outflow; lower panels) of J0954$+$47. In the upper panels, the small diamonds with error bars are the \siiietext\ outflows in star-forming galaxies from \citet{Chisholm2015}. In these two panels, the values of \vwu\ are all measured from the overall absorption features. In the bottom panels, the outflows in the CLASSY sample from \citet{Xu2022} are shown in small diamonds with error bars. For each object, \vwu\ is the median value of all absorption features detected. In these two panels, \vwu\ is measured from the highest velocity component of the best-fit model. The black dashed lines in the two right panels indicate the expected escape velocity $v_{esc} \simeq 3v_{circ}$. In the two panels on the left, the data points are colored by stellar mass. In the two panels on the right, the data points are colored by SFR (star-forming sources) or the upper limits on SFR (J0954$+$47).}
\label{fig:sfr}
\end{figure*}

\subsubsection{Comparison with Starburst-driven Outflows} \label{05323}

Star-formation-driven outflow is a popular feedback mechanism in dwarf galaxies. To understand the nature and impact of the fast outflow in J0954$+$47, we compare its properties with those of star-formation-driven outflows in low-z galaxies observed by \textit{HST}/COS.

As shown in Fig. \ref{fig:sfr}, we plot the outflow velocities as functions of SFR\footnote{The upper limit on SFR for J0954$+$47 is derived from the \oiiab\ luminosity \citep[based on][see L20 for details]{Ho2005}, while the SFR of star-forming galaxies from \citet{Xu2022} are obtained from \citet{CLASSY}, which are based on UV$+$optical spectral energy distribution (SED) fitting, and the SFR of star-forming galaxies from \citet{Chisholm2015} are obtained adopting both IR and UV data. There may be systematic differences between these three types of SFR measurements, but we do not expect these discrepancies to be large enough to change our results qualitatively.} and circular velocities for both J0954$+$47 and those from \citet{Chisholm2015} or \citet{Xu2022}. {We only include sources with UV radii $r_{50}< 1.5$\arcsec\ from \citet[][i.e., red filled circles in their Fig. 9]{Xu2022}, as the UV light of these sources lies within the \textit{HST}/COS aperture, which allows for reliable measurements of their global outflow properties.} In \citet{Chisholm2015}, the outflow velocity is measured from the overall best-fit \siiic\ profile. In \citet{Xu2022}, the outflow velocity is measured as the median \vwu\ of the blueshifted components in the best-fit model for all transitions detected, including \oiuv, \cii, \siiiabcde, \siiii, and \siiv. For a fair comparison, therefore, we show two outflow velocities for J0954$+$47: the upper panels show \vwu\ of the overall absorption line profile as in \citet{Chisholm2015}, and the bottom panels show \vwu\ of the high-v outflow, similar to that defined in \citet{Xu2022}. In all panels, the data points are colored according to their SFR or stellar mass.

With a SFR of $<$ 0.3 \msunyr, the outflow in J0954$+$47 is much faster than those in star-forming galaxies with similar SFR from both \citet{Chisholm2015} and \citet{Xu2022}. The presence of the AGN in this dwarf galaxy appears to boost the outflow velocity. {This is in agreement with the results in \citet{Aravindan2023} where they find a larger outflow velocity in dwarf galaxies with AGN than star-forming dwarf galaxies based on an integral field spectroscopy study of emission lines. This is also consistent with the results from the simulations in \citet{Koudmani2019} where the presence of AGN feedback in dwarf galaxies leads to higher outflow velocity than pure stellar feedback.}

Given the circular velocity (\vcirc) of $\sim$ 70 \kms\ (based on the stellar mass of J0954$+$47), the outflow in J0954$+$47 is also faster than those star-formation-driven ones from \citet{Chisholm2015} with similar circular velocities and is among the fastest outflow seen among the star-formation-driven ones from \citet{Xu2022}. This suggests that, for a given gravitational potential, AGN can at least drive an outflow as fast as those seen in dwarf galaxies with extreme starbursts. 

\begin{figure*}
\begin{minipage}[t]{0.33\textwidth}
\includegraphics[width=\textwidth]{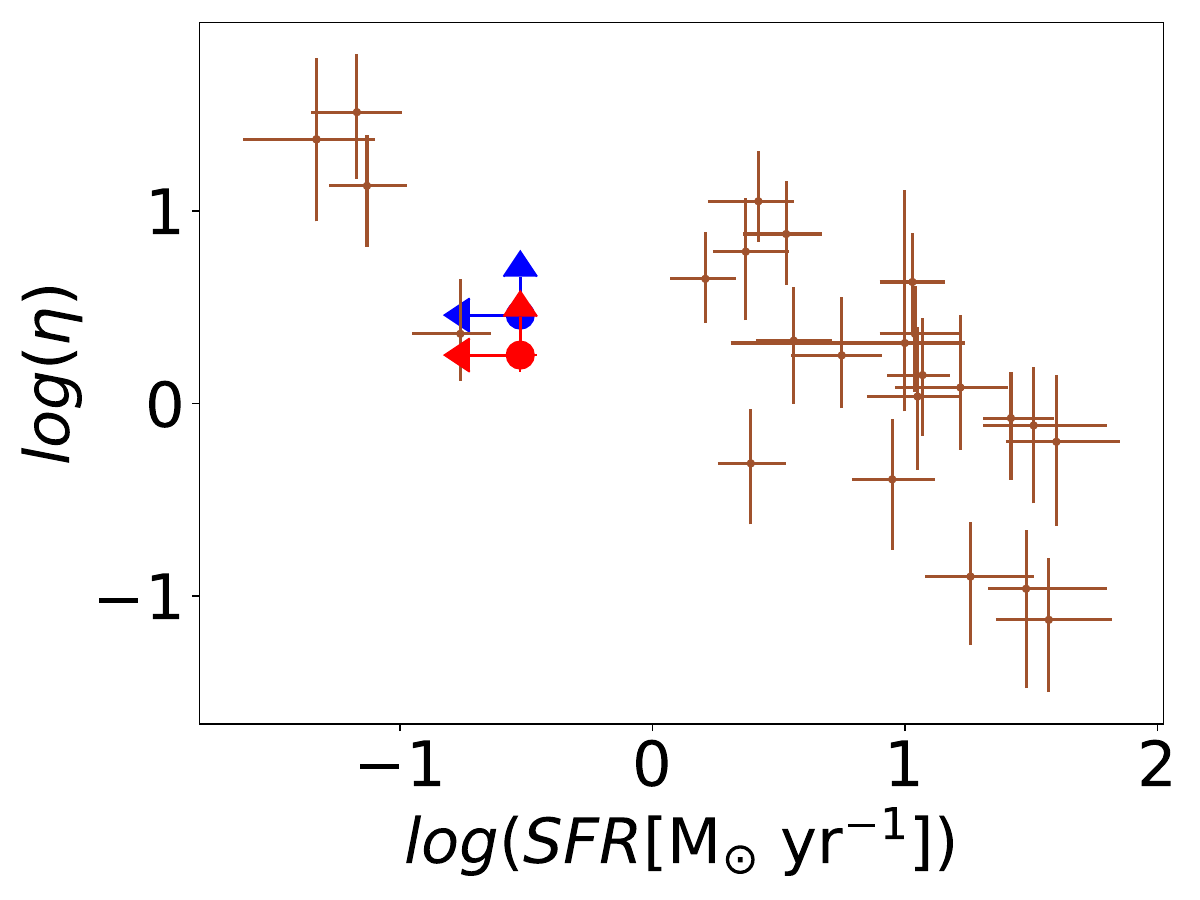}
\end{minipage}
\begin{minipage}[t]{0.33\textwidth}
\includegraphics[width=\textwidth]{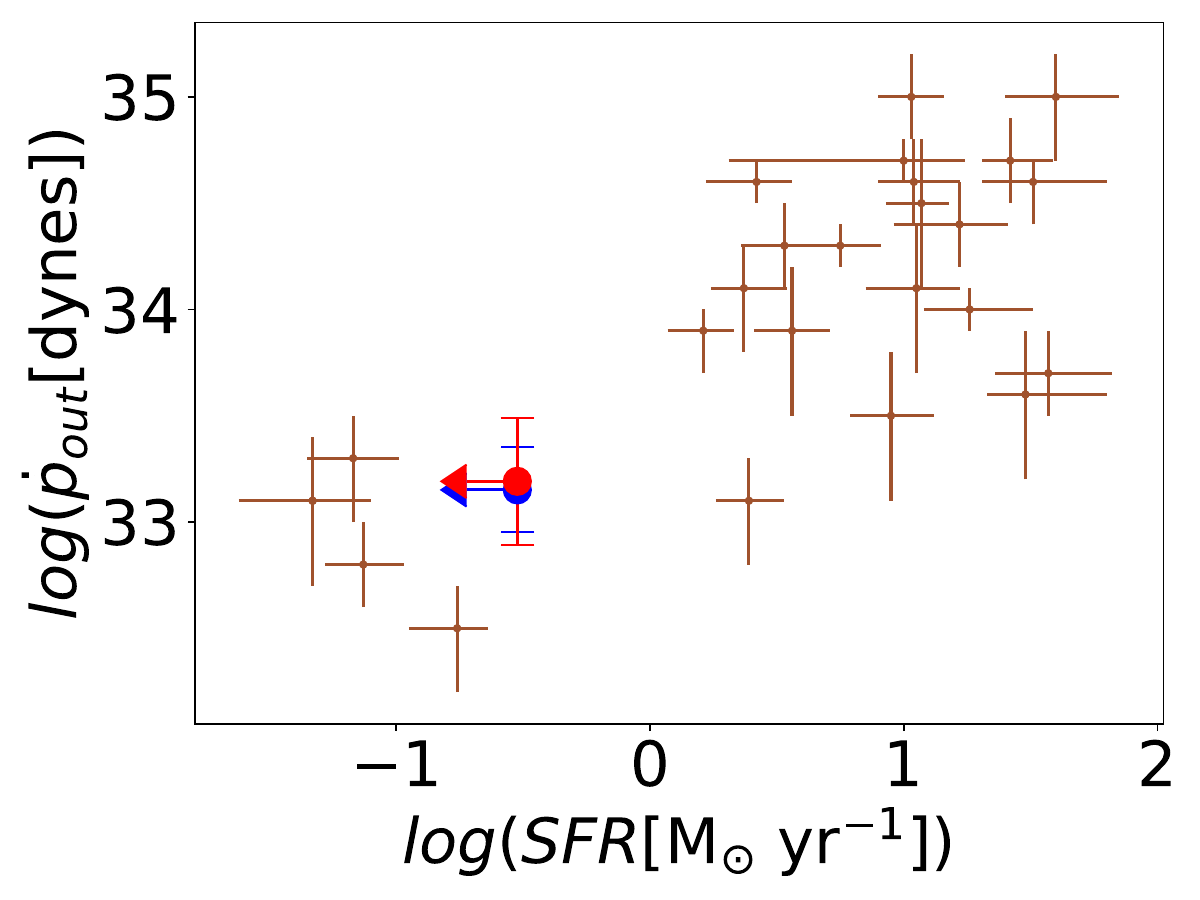}
\end{minipage}
\begin{minipage}[t]{0.33\textwidth}
\includegraphics[width=\textwidth]{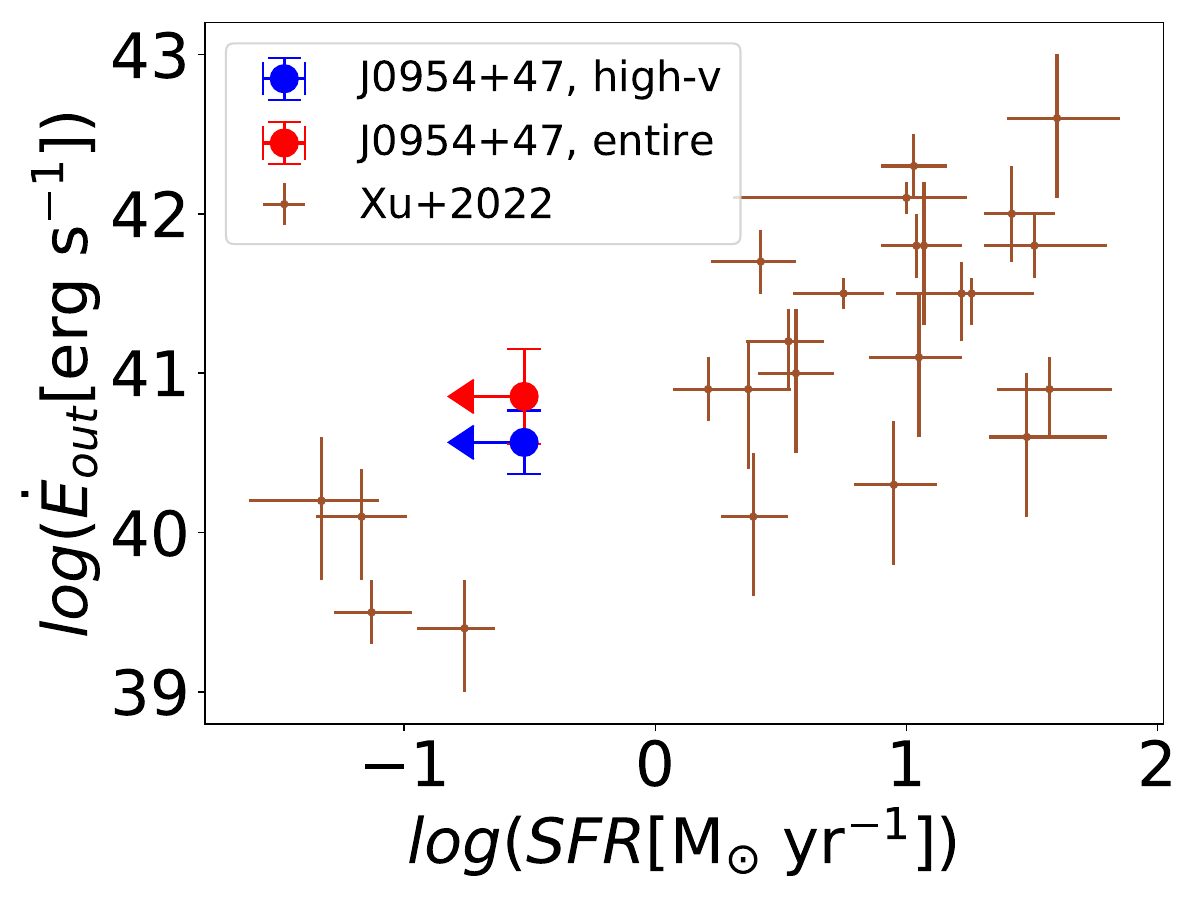}
\end{minipage}

\begin{minipage}[t]{0.33\textwidth}
\includegraphics[width=\textwidth]{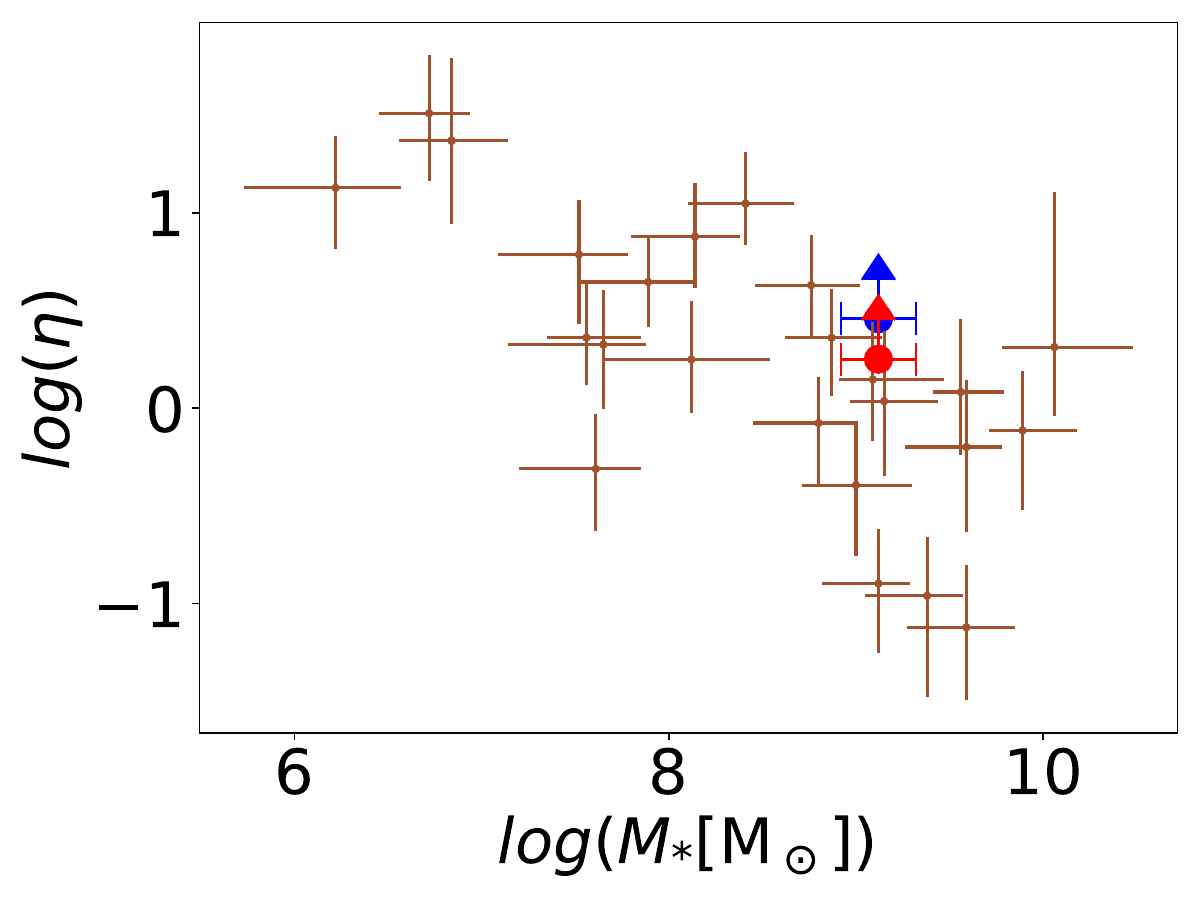}
\end{minipage}
\begin{minipage}[t]{0.33\textwidth}
\includegraphics[width=\textwidth]{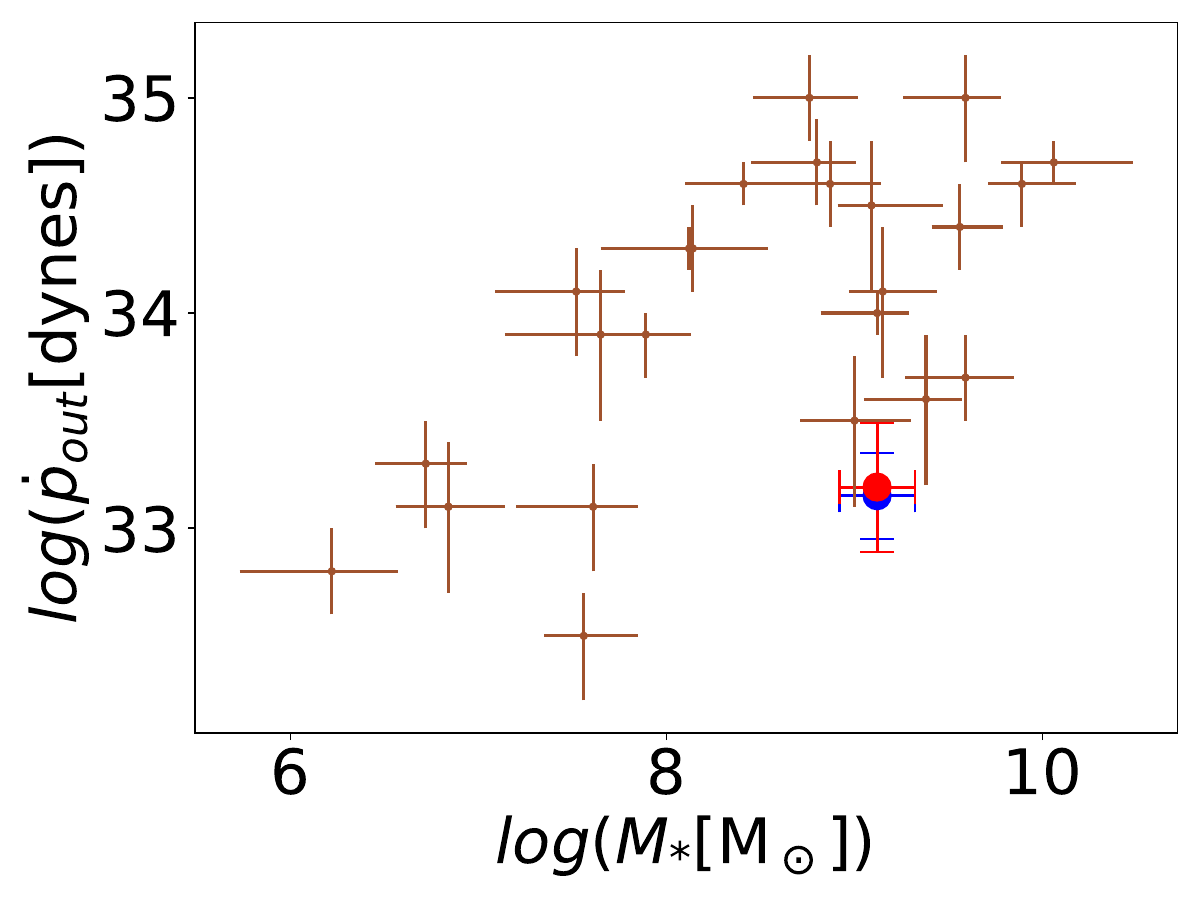}
\end{minipage}
\begin{minipage}[t]{0.33\textwidth}
\includegraphics[width=\textwidth]{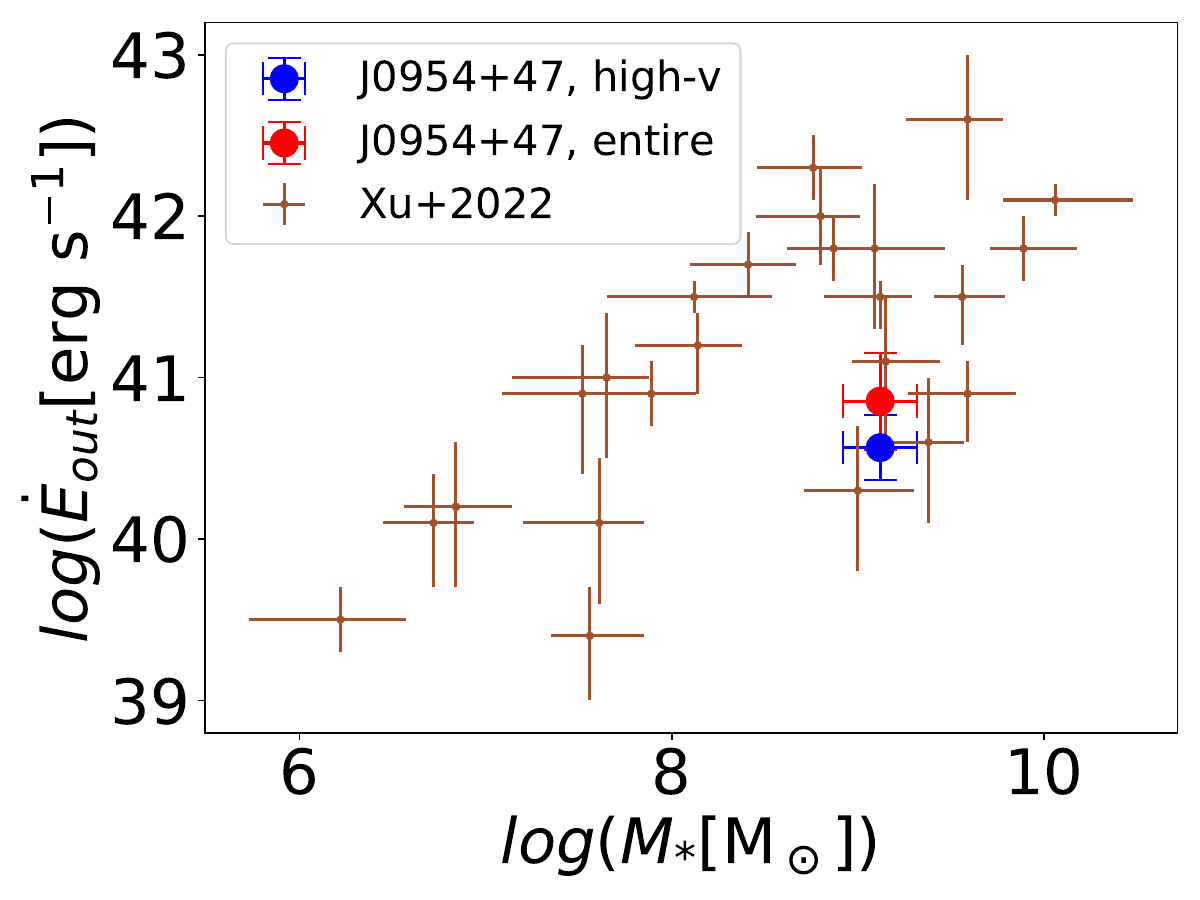}
\end{minipage}
\caption{{Mass loading fatcor $\eta$ {(left column)}, momentum outflow rate {(middle column)} and kinetic energy outflow rate {(right column)} as a function of SFR (top row) and stellar mass (bottom row) for J0954$+$47 (red and blue filled circles) and star-forming galaxies from \citet{Xu2022} (brown crosses). The red and blue filled circles represents the average
value for the entire outflow and the value for the high-v outflow in J0954$+$47, respectively.}}
\label{fig:compare}
\end{figure*}

{Moreover, in Fig. \ref{fig:compare}, we further compare the mass loading factor (defined as the ratio of mass outflow rate to SFR), momentum outflow rate and kinetic energy outflow rate of the outflow in J0954$+$47 with those of star formation-driven outflows in low-z galaxies, taking sources from \citet{Xu2022} as an example. }

{For a given SFR, the mass loading factor and momentum outflow rate of J0954$+$47 are comparable to those of star-forming galaxies, whereas the kinetic energy outflow rate is larger than those of star-forming galaxies by $\sim$1 dex. These are again largely consistent with the results from the 
IFS study presented in \citet{Aravindan2023} on optical emission lines, where they find higher mass, momentum and kinetic energy outflow rates in dwarf galaxies with AGN than star-forming ones, and the difference increases from the mass rate to momentum rate and is the largest in the energy rate. Our results also broadly agree with the results from the simulations in \citet[][]{Koudmani2019}: while they find no significant difference in mass loading factors between the simulation run with both AGN and supernovae feedback and the one with only supernovae feedback, the thermal energy outflow rate in the former is significantly larger than that in the latter by up to  $\sim$5 dex. These results are in concordance with the picture that, at a given SFR, compared to stellar feedback, the AGN feedback deposits more energy to heat the ISM and thus prevents the cooling of gas while having a similar impact on expelling the gas out. However, this is based on only one dwarf galaxy with AGN and a larger sample is needed before any robust conclusion can be drawn.}

{For a given stellar mass, while the mass loading factor of J0954$+$47 is still comparable to the maximum values in star-forming galaxies, the momentum and kinetic energy outflow rates of J0954$+$47 are close to the lowest values observed in the star-forming galaxies. This is likely caused by the extreme starburst activities in the sources from \citet{Xu2022} that are not typical for z$\sim$0 dwarf galaxies like J0954$+$47. The SFR of galaxies from \citet{Xu2022} with similar stellar mass to J0954$+$47 are $\sim$1--2 dex larger than both the upper limit of SFR in J0954$+$47 and the average SFR at z$\sim$0 \citep[See the left panel of Fig. 3 in][]{Berg2022}. Instead, the range of SFR in the \citet{Xu2022} sample is comparable to those star-forming galaxies at z$\sim$1--2.5, which may represent the most extreme cases of stellar feedback, rarely seen in z$\sim$0 star-forming dwarf galaxies. The much higher SFR in their sample thus allows for much higher energy output to drive outflows more powerful than the presumably AGN-driven outflow in J0954$+$47. Therefore, at a given stellar mass, the most extreme stellar feedback may still play a more important role than AGN feedback in dwarf galaxies, if J0954$+$47 is a representative case for AGN feedback in dwarf galaxies. However, this is again based on only one dwarf galaxy with AGN, and it would be interesting to further look for more extreme AGN activities in dwarf galaxies at higher redshift that may provide stronger feedback.}

\begin{figure}
\plotone{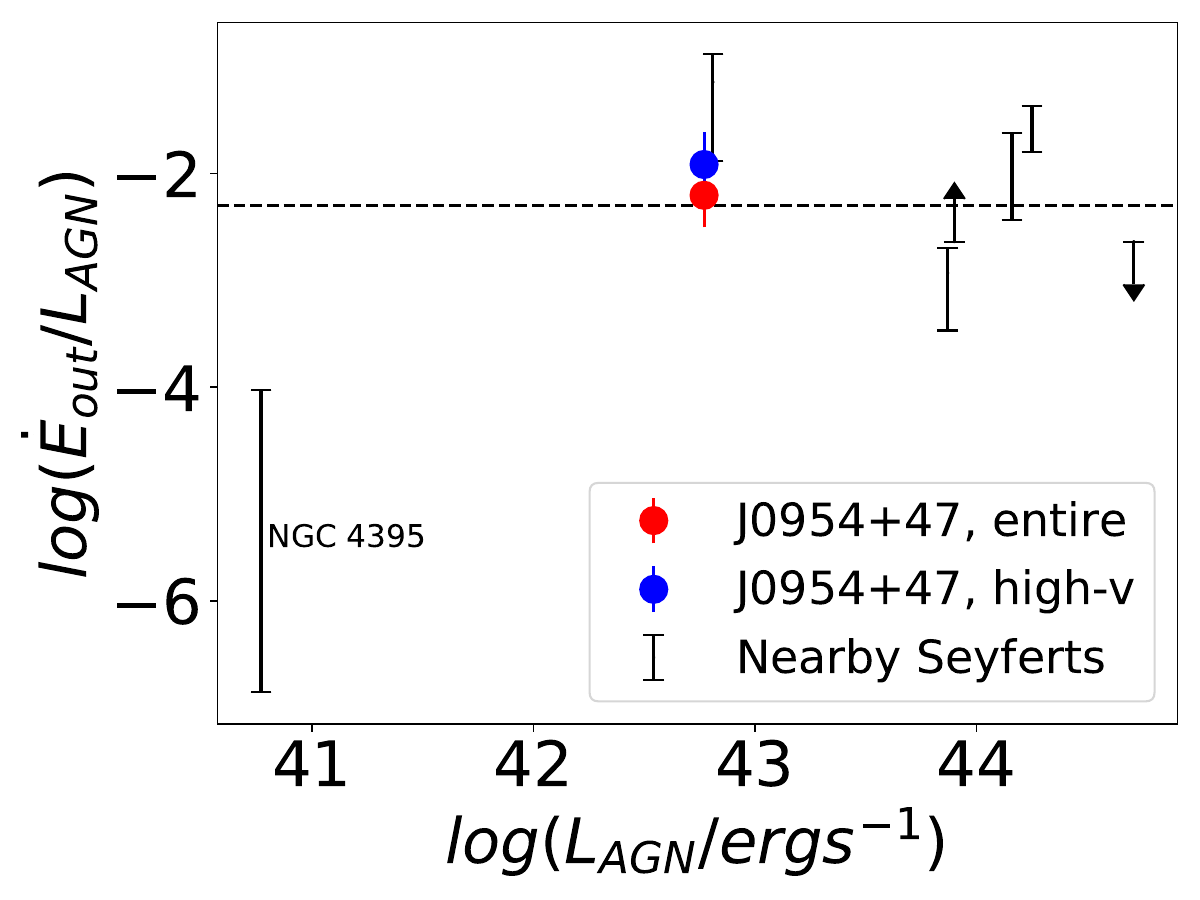}
\caption[]{Ratio of kinetic energy outflow rate to AGN luminosity as a function of AGN luminosity for the average value of the entire outflow (red) and for the high-v outflow (blue) of J0954$+$47. Also shown in black are the X-ray and FUV absorber-traced outflows in nearby Seyferts from \citet{Crenshaw2012}. Note that in the bottom left corner is the outflow in the archetype dwarf Seyfert 1, NGC~4395. The horizontal dashed line at 0.5\% is the prediction for effective AGN feedback from \citet{HopkinsElvis2010}.}
\label{fig:LAGN}
\end{figure}

\begin{figure*}
\plottwo{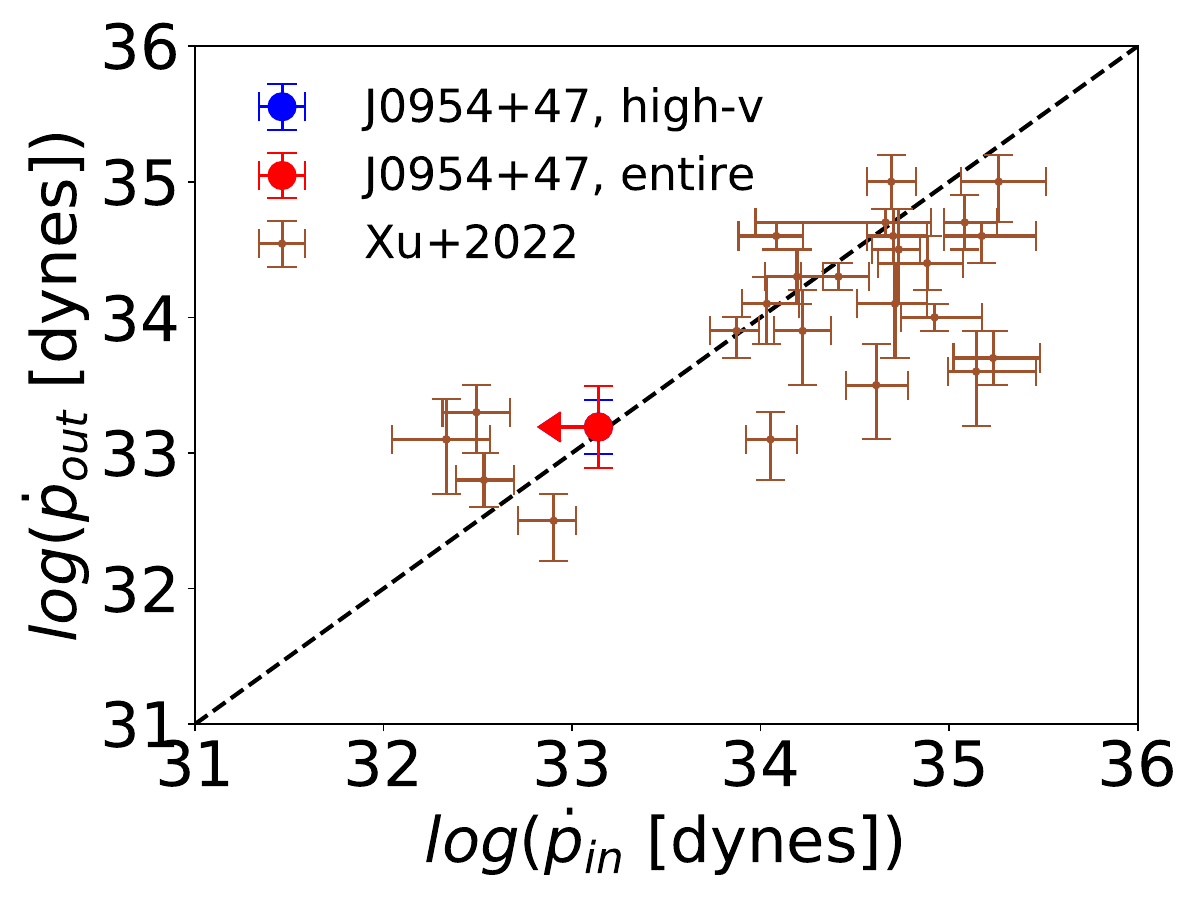}{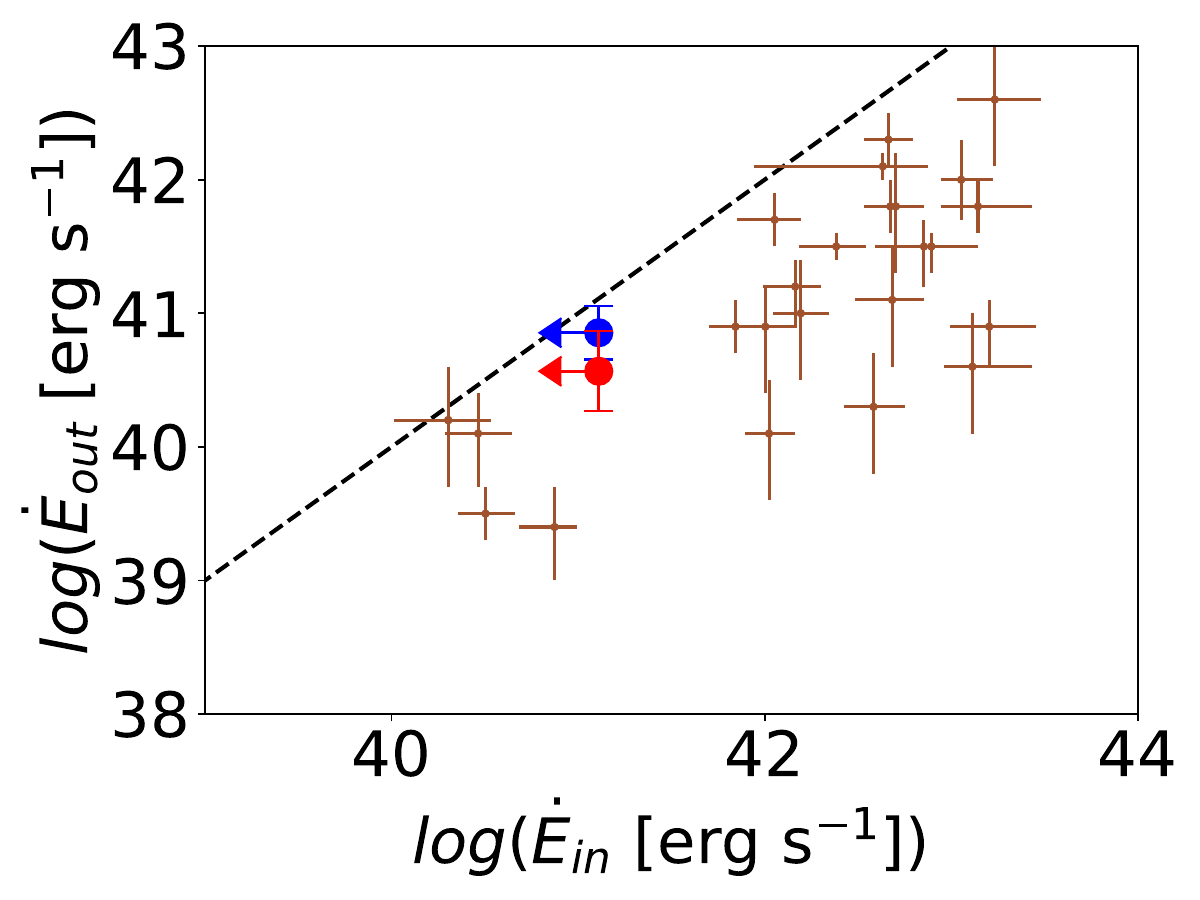}
\caption{{Momentum (left) and kinetic energy (right) outflow rates versus expected momentum (left) and energy (right) injection rates from the star formation activity (see Section \ref{05324}) for the average value of the entire outflow (red) and high-v outflow (blue) in J0954$+$47, and those in low-z star-forming galaxies from \citet{Xu2022} (brown crosses). The dashed lines in both panels indicate the 1:1 equality line.}}
\label{fig:pout}
\end{figure*}

\subsubsection{Energy Source of the Outflow} \label{05324}

{In this section, we explore the possible energy source for the outflow in J0954$+$47.
}

First, the AGN is powerful enough to easily drive the outflow, since the ratio of the kinetic energy outflow rate to AGN luminosity, \dedtlagn, is $\sim$0.6\% (the average for the entire outflow) or $\sim$1.2\% (for the high-v outflow), comparable to low-z AGN with outflow measured with the same technique (Fig. \ref{fig:LAGN}; the \lagn\ value adopted here is derived based on the \textit{HST}/COS data and the AGN template from CLOUDY. See Appendix \ref{A1} for more details). 

{Second, as shown in Fig. \ref{fig:sfr} and stated at the end of Section \ref{05323}, the outflow is faster than those of outflows in dwarf galaxies with similar SFR, implying that the presence of AGN is likely required to reach such a high outflow velocity, which is also consistent with simulations \citep[e.g.,][]{Koudmani2019}.}

{Finally, it is challenging but still possible in principle for star formation activity alone to inject enough momentum and energy to the outflow. The total momentum injection rate supplied by the star formation activity, $\dot{p}_{in} = 4.6 \times 10^{33} SFR~[\msun\ yr^{-1}]$ dynes, which is the sum of the hot wind fluid driven by thermalized ejecta of massive stars \citep{Chevalier1985} and radiation pressure \citep{Murray2005}. Similarly, the total energy injection rate supplied by the star formation activity is $\dot{E}_{in} = 4.3 \times 10^{41} SFR~[\msun\ yr^{-1}]$ erg s$^{-1}$.  A SFR $<0.3$ \msunyr\ translates into a momentum injection rate of $<1.4 \times 10^{33}$ dynes and an energy injection rate of $<1.3 \times 10^{41}$ erg s$^{-1}$. Given the measured momentum (kinetic energy) outflow rate $\dot{p}_{out}$ $\simeq 1.6^{+1.6}_{-0.8}\times 10^{33}$ dynes ($\dot{E}_{out}$ $\simeq 7.9^{+7.9}_{-4.0} \times 10^{40}$ erg s$^{-1}$), the star formation activity is capable of driving the outflow if it can transfer the momentum and energy to the outflow at high efficiencies ($\gtrsim$60\% -- 230\% for momentum and $\gtrsim$30\% -- 120\% for kinetic energy). While possible, these are not typical for nearby star-forming galaxies: as shown in Fig. \ref{fig:pout}, for the majority of the star-forming galaxies in \citet{Xu2022}, the ratios of momentum (kinetic energy) outflow rates to the momentum (energy) injection rates are lower than those of J0954$+$47.
Moreover, the momentum and energy injection rates from star formation activity adopted above are uncertain. For example, if adopting the estimates from \citet{Rupke2023} and scaling the SFR that for our object, we obtain a $\dot{p}_{in} < 4 \times 10^{32}$ dynes and $\dot{E}_{in} < 8 \times 10^{40}$ erg s$^{-1}$ for SFR $<$ 0.3 \msunyr. The required efficiencies increase to $\gtrsim$210\% -- 810\% for momentum and $\gtrsim$50\% -- 200\% for energy, making it even more difficult for the star formation activity to drive the outflow. }

{Overall, AGN is likely the dominating energy source for the outflow. Nevertheless, we cannot formally rule out the possibility that star formation activity may also help launch the outflow in J0954$+$47.}

\subsubsection{The Impact of the Outflow} \label{05325}

Given the measured large velocity of the outflow in J0954$+$47, some of the outflowing gas may escape the host galaxy altogether.
We estimate the escape velocity as $v_{esc}^2 = 2v_{circ}^2[1+ln(r_s/R)]$ where the mass distribution is modeled as a truncated isothermal sphere: $r_s$ is the radius of the sphere, and R is the radius where the \vesc\ is calculated as $v_{esc} \simeq 3v_{circ}$ with $r_s/R \simeq 33$ as adopted in literature \citep[e.g., see][for a brief discussion]{Veilleux2020}. The average velocity of the entire outflow ($\sim$260 \kms) and velocity of the high-v outflow ($\sim$460 \kms) in J0954$+$47 are comparable to, or significantly larger than, \vesc\ of the host galaxy ($\sim$220 \kms). A large portion ($\sim$50\% in terms of absorption line equivalent width) of the ionized gas traced by the absorption may escape the system and help enrich the circumgalactic medium.

The outflow also carries significant amounts of mass, momentum, and kinetic energy. We obtain a mass loading factor (i.e., mass outflow rate over SFR) of $\gtrsim$3.4 ($\gtrsim$1.8) for the average of the entire outflow (the high-v outflow), suggesting that more gas is carried away by the outflow than that fueling star formation. {As shown in Fig. \ref{fig:LAGN},} the kinetic energy outflow rate to AGN luminosity, \dedtlagn, is $\sim$0.6\% ($\sim$1.2\%) for the average of the entire outflow (the high-v outflow), which is comparable to (larger than) the expectation from one representative simulation of AGN feedback \citep[e.g., ][]{HopkinsElvis2010} that only require $\sim$0.5\% of the AGN bolometric luminosity to be injected into the outflow for effective AGN feedback.\footnote{According to \citet{HopkinsElvis2010}, AGN feedback will be effective if $\eta \dot{m} \frac{M_{BH}}{<M_{BH}>} \sim 0.05f_{hot}$, where $\eta = \dot{E}_{out}/L_{AGN}$, $\dot{m}$ is the Eddington ratio, ${M_{BH}}$ and ${<M_{BH}>}$ are the black hole mass and the expectation from the M-$\sigma$ relation, and $f_{hot}$ is the hot gas fraction which is typically $\sim$0.1 in ISM. If J0954$+$47 follows M-$\sigma$ relation ($\frac{M_{BH}}{<M_{BH}>}=1$), then $\dot{m} \sim$1. Therefore, it leads to $\dot{E}_{out}/L_{AGN} \sim 0.5\%$ for an effective feedback.\\ In addition, the prediction from such simulations are the kinetic coupling efficiency, the fraction of the AGN bolometric luminosity coupled to gas near the black hole. This is always larger than the observed \dedtlagn: Even for an energy-driven large-scale outflow, only a portion of the original nuclear wind power ends up in kinetic form, where the rest of the energy may be used up in doing work against the gravitational potential and ambient pressure along the way \citep{Harrison2018}.} Therefore, the outflow in J0954$+$47 is likely powerful enough to provide significant negative feedback to its dwarf host galaxy. Interestingly, the \textit{HST} image of this object (Fig. \ref{fig:morphology}b) shows no evidence of spiral arms and suggests no significant on-going star formation. The color of this source based on SDSS images is also green ($u-r \sim 2$), suggesting that it is transitioning from a blue star-forming galaxy to a red, passive galaxy \citep[][]{ManzanoKing2019}. These properties are apparently consistent with the star formation quenching caused by the negative feedback via the outflow.

Note that the kinetic energy outflow rate derived based on absorption lines here is $\sim$10$\times$ larger than that obtained from previous \oiii\ emission line studies \citep[$log(\dot{E}_{out}) \sim 39.6$ if the outflow is spatially resolved;][]{Liu2020}. This discrepancy may be largely due to the fact that \oiiitext\ emission line misses the lower density and/or lower-ionization outflowing gas which is more easily probed by the absorption features examined in this paper. This emphasizes the importance of probing outflows with different emission/absorption tracers to acquire a complete census on the outflow energetics.

\begin{figure*}
\centering
 \begin{minipage}[t]{0.48\textwidth}
\includegraphics[width=\textwidth]{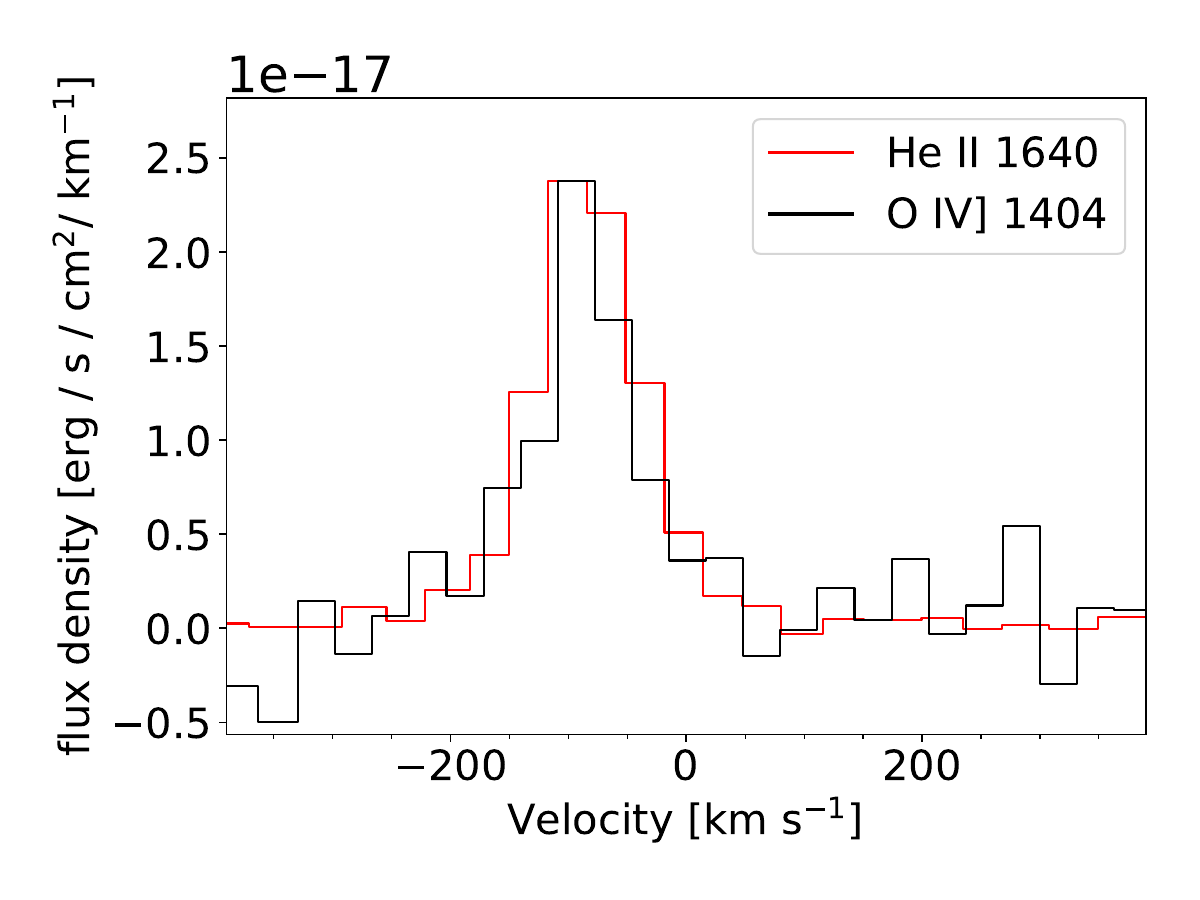}
\end{minipage}
\begin{minipage}[t]{0.48\textwidth}
\includegraphics[width=\textwidth]{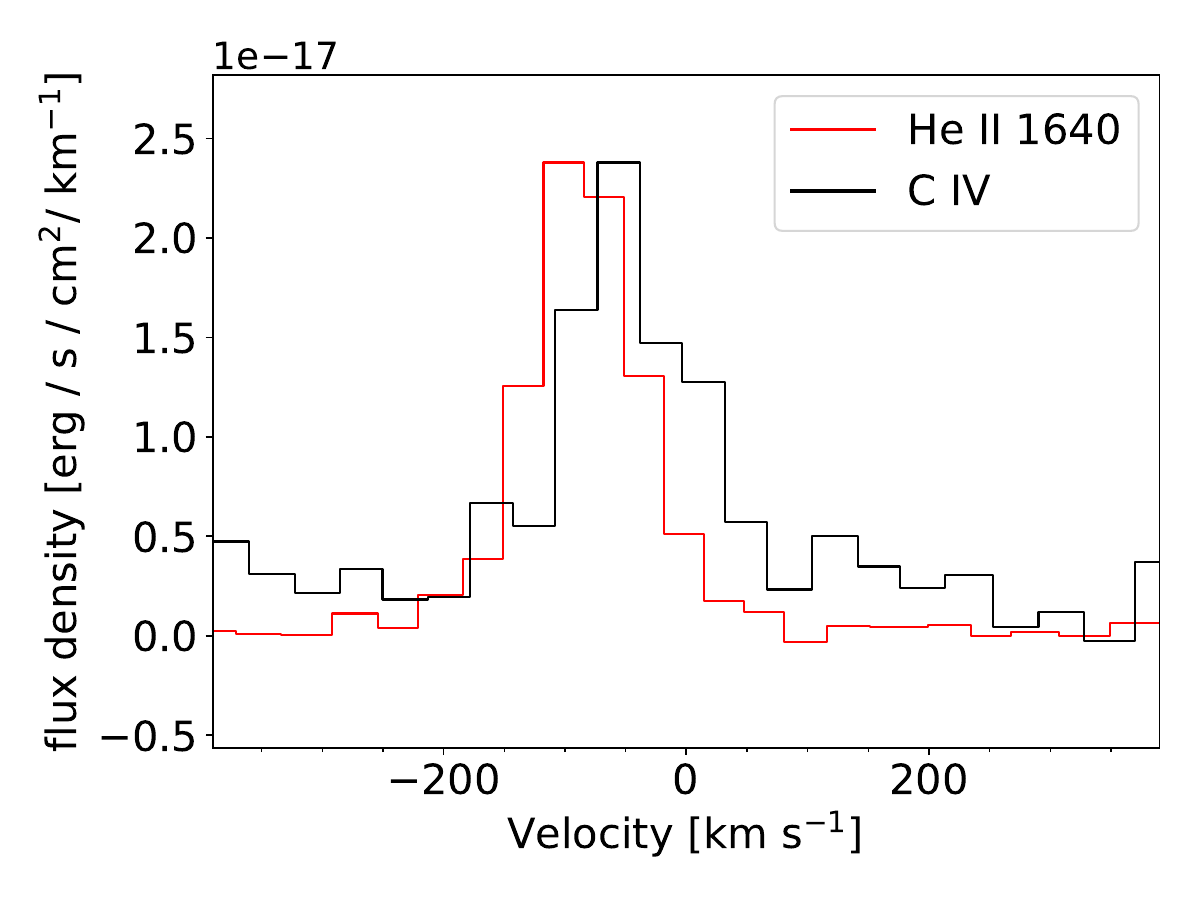}
\end{minipage}

\centering
 \begin{minipage}[t]{0.48\textwidth}
\includegraphics[width=\textwidth]{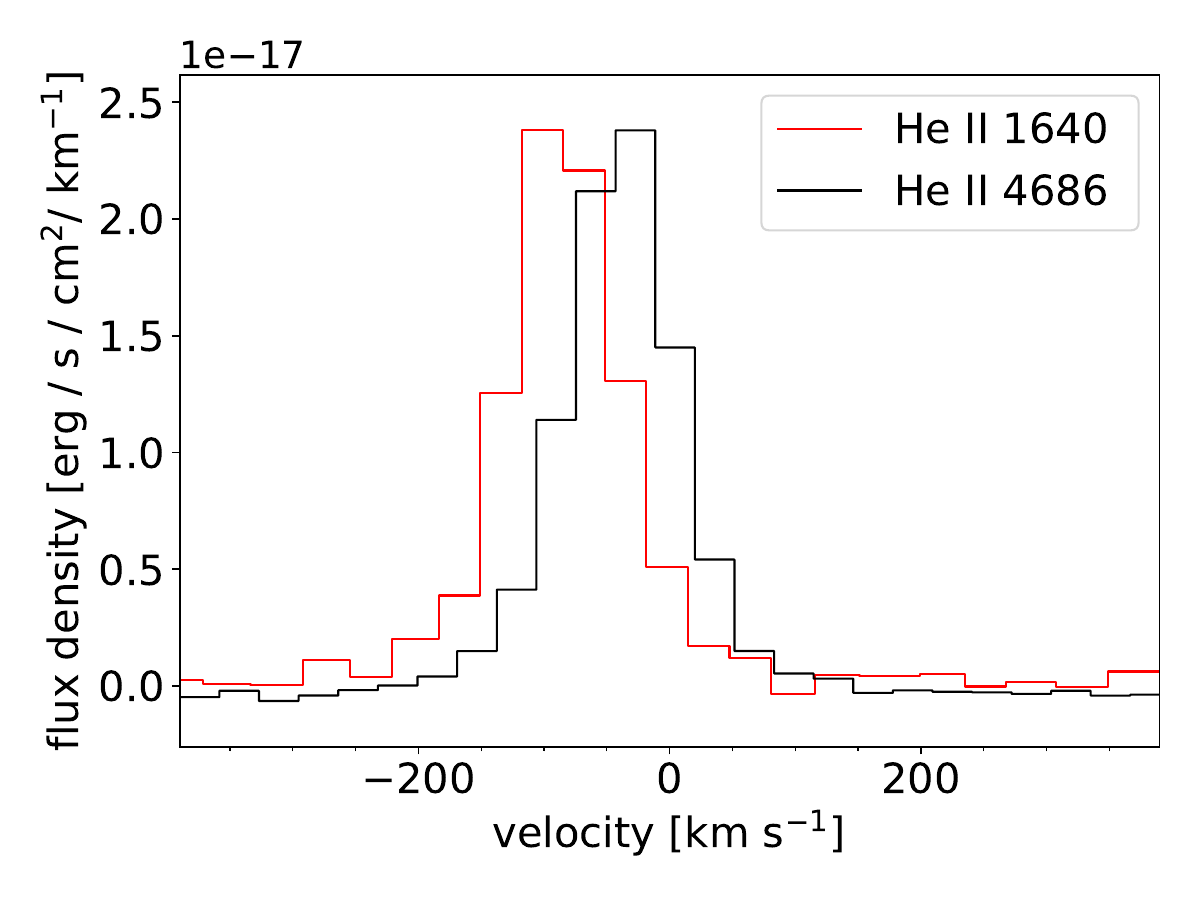}
\end{minipage}
 \begin{minipage}[t]{0.48\textwidth}
\includegraphics[width=\textwidth]{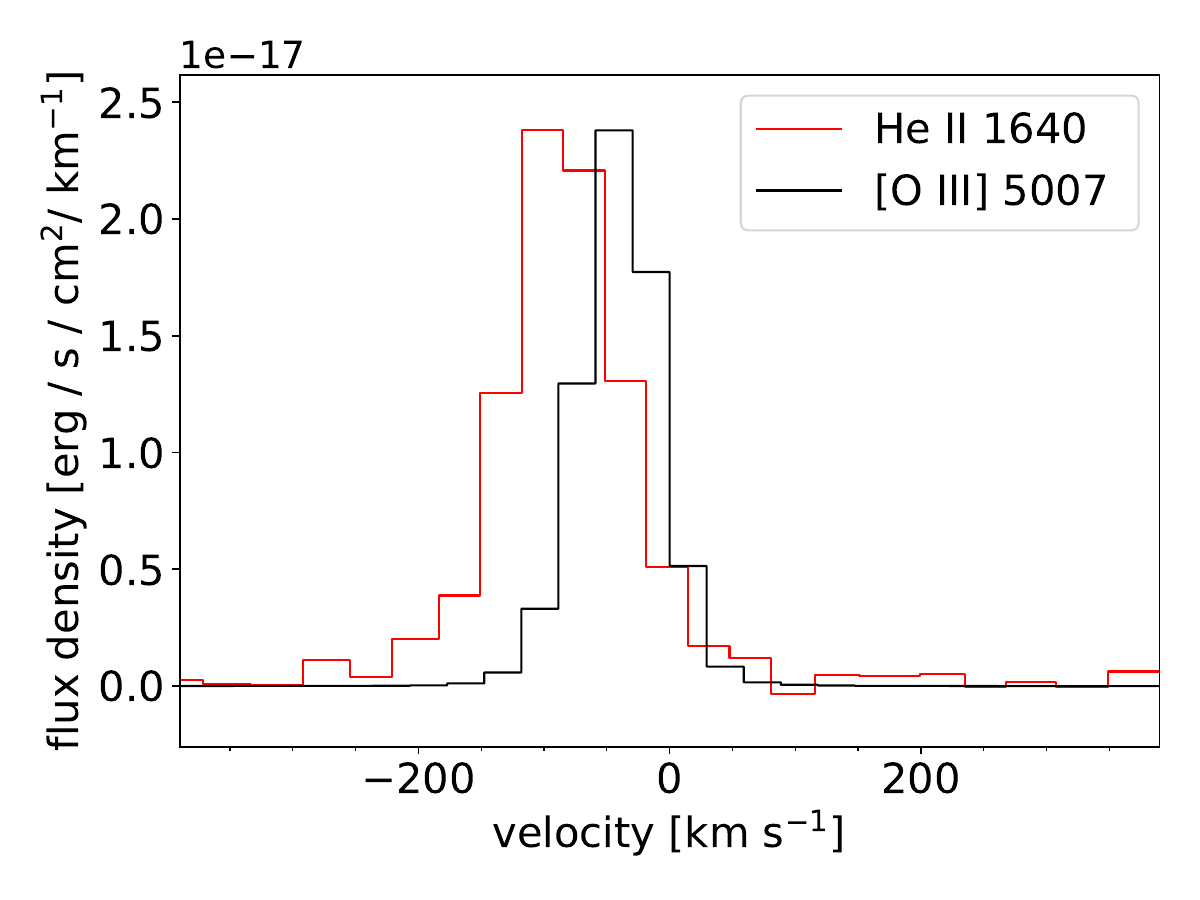}
\end{minipage}
\caption{\heii\ emission line profile compared with \oivb, \civb, \heiiopt, and \oiii\ emission line profiles  for J1009$+$26, respectively. All spectra are continuum subtracted. The \heii\ emission line is in its original flux scale while all other emission lines are re-scaled to have the same peak flux density as that of \heii. The systemic velocity is determined from the stellar kinematics based on the IFS data from L20 as listed in Table \ref{tab:targets}.}
\label{fig:J1009He2}
\end{figure*}

\section{\heii\ Emission} \label{054}

\subsection{Detection and Basic Properties} \label{0541}

A strong \heii\ emission line is detected in two objects, J0954$+$47 (Appendix \ref{A2})) and J1009$+$26 (Fig.\ \ref{fig:J1009He2} and Appendix \ref{A2}), consistent with the AGN nature of these two objects, given the high ionization potential of He$^{+}$ ($\sim$54.4 eV). The equivalent widths of \heiitext\ are 2.71$\pm{0.14}$ \AA\ and 7.00$\pm{0.29}$ \AA, respectively, again broadly consistent with the high values seen in AGN \citep[with a median \heiitext\ equivalent width of $\sim$8 \AA; e.g.][]{Hainline2011,Cassata2013}. The \heiitext/\civtext\ flux ratios of the two objects are $\sim$0.8 and $\sim$1.7, which fall in the range ($\sim$0.002$-$3) predicted by AGN photoionization models \citep[e.g.][]{Feltre2016}. In
 addition, the \heiitext/\civtext\ flux ratios in the two objects are also broadly consistent with values typically measured in Type 2 quasar/AGN \citep[$\sim$ 0.7; e.g.][]{McCarthy1993,CorbinBoroson1996, Humphrey2008,Matsuoka2009}.

The \heiitext\ lines in these two sources are narrow, with FWHM of 120$\pm{10}$ \kms\ and 110$\pm{10}$ \kms\ for J0954$+$47 and J1009$+$26, respectively. Therefore, these are not the broad \heiitext\ features observed in Wolf-Rayet stars/galaxies \citep[e.g.][]{WRstar}.  
Moreover, to our knowledge, for star-forming galaxies in the nearby universe, narrow nebular \heiitext\ emission lines have only been observed in metal-poor sources \citep[e.g.][]{Berg2016,Senchyna2017}, while the metallicity of our two objects, J0954$+$47 and J1009$+$26, are both close to solar values based on the modeling of optical line ratios from L20. Additionally, the equivalent widths of those \heiitext\ lines in nearby star-forming galaxies (up to $\sim$1.7 \AA\ as reported in those two references) are lower than those of J0954$+$47 and J1009$+$26.

Overall, the \heii\ emission lines detected in our sample are consistent with an AGN origin. A more detailed analysis that takes into account all FUV, optical, and near-IR emission lines with various ionization potentials will be the subject of a future paper. The hard ionization spectra from dwarf galaxies with AGN like our objects may lead to high escape fractions of ionizing photons, which, if present in abundance in the early universe, may contribute to cosmic reionization \citep[e.g.][]{Madau2015,Robertson2015}. Interestingly, the recent discovery of a population of faint AGN at z $\sim$ 4--10 \citep[e.g.][]{Matthee2023,Harikane2023,Maiolino2023,Ubler2023} suggests that such dwarf AGN may indeed be more common at the epoch of reionization than previously expected.

\subsection{Blueshifted \heii\ Emission Tracing A Highly-ionized Outflow in J1009$+$26?} \label{0542}

An interesting characteristic of the \heii\ emission line in J1009$+$26 is its blueshift. As shown in Fig. \ref{fig:J1009He2}, \vwu\ of \heii\ is $-$70$\pm{5}$ \kms\ with respect to the systemic velocity based on the stellar kinematics measured in L20. As a comparison, \vwu\ of \oivb, \heiiopt, \civ, and \oiii\ are $-$65$\pm{5}$ \kms, $-$35$\pm{5}$ \kms, $-$30$\pm{15}$ \kms, and $-$20$\pm{15}$ \kms, respectively. The ionization potentials (IP) for the corresponding ions are 54.9 eV, 54.4 eV, 47.9 eV, 35.1 eV, respectively. \oivtext\ and \heii\ emission lines are the most blueshifted ones among these lines.

One possible origin of the blueshifted \heii\ emission line is a highly-ionized AGN wind. As mentioned in Section \ref{0541}, \heii\ is likely originating from the highly ionized gas in the vicinity of the AGN. \oivtext\ lines likely share a similar origin to \heii\ based on their similar IP. As a comparison, \civtext\ and \oiiitext\ may have a significant contribution from the ISM further away from the AGN given their lower IP. This trend of increasing outflow velocity in higher ionization species in our objects is also suggested by \citet{Bohn2021} where the outflow velocity measured from the \sivi\ (IP$\simeq$166.8 eV) is faster than that from the \oiiitext\ line. These results are all consistent with a ionization-stratified outflow model for AGN-driven outflow in more luminous AGN \citep[e.g.][]{Veilleux1991,Spoon2009}, namely the outflow velocity decreases with distance to the central ionizing source and so does the ionization. A more detailed discussion on this combining all multi-wavelength data is beyond the scope of this paper and thus saved for a future work. 

Therefore, the blueshift of \heii\ and \oivtext\ lines are naturally caused by the motion of the highly-ionized AGN wind towards the observer, whereas the blueshift of \civtext\ and \oiiitext\ are smaller due to the contamination from the nebular emission of the relatively more kinematically quiescent ISM. 

Moreover, the blueshift of \heii\ with respect to \heiiopt\ might be caused by the heavier attenuation in the FUV than in the optical: The red emission line wing of \heii\ originating from the material on the far-side of the AGN is attenuated by gas/dust more severely than that of \heiiopt. The \heii\ is thus more blueshifted overall due to the much weaker red emission line wing. Likewise, the velocity offset between the \heii\ and \oiiitext\ emission lines may also partially be caused by this effect, in addition to the contamination from the ISM in the latter as mentioned above.

\section{Summary}

In this paper, we present the results from a pilot \textit{HST}/COS spectroscopy program to examine three dwarf galaxies with primarily AGN-driven outflows from the sample studied in \citet{Liu2020}. The main results are summarized as follows:

\begin{itemize}

\item 
Blueshifted absorption features are clearly detected in two objects (J0906$+$56 and J0954$+$47) and tentatively detected in one object (J1009$+$26), tracing outflows in them. 

\item 
In J0954$+$47, fast outflow is detected in multiple transitions including \ciitext, \civtext, \siiietext, and \siivtext. The outflow velocity is $\sim-$460 \kms\ for the high velocity component or $\sim-$230 \kms\ averaged over the entire absorption feature. In addition, much broader absorption features ($\sim$4000 \kms) are also seen for \civtext\ and \siivtext, which may have a similar origin to the broad absorption lines in quasars.

\item 
In J0954$+$47, the radius of the outflow, based on the results from photoionization modeling with CLOUDY, is estimated to be $\sim$0.5 kpc. The velocity of this outflow exceeds the escape velocity of the system, suggesting that a large fraction of the outflowing gas may escape the system. This outflow may help deposit energy into the circumgalactic medium and enrich it with metals.

{
\item
We compare the outflow properties of J0954$+$47 to those of star formation driven outflows from \citet{Chisholm2015} and \citet{Xu2022}: At a given SFR, this outflow is much faster and possesses a higher kinetic energy outflow rate than star-forming galaxies. The mass loading factors and momentum outflow rates are comparable between the two. This is consistent with the scenario that at a given SFR, compared to stellar feedback, the AGN feedback deposits more energy to heat the ISM and thus prevents the cooling of gas while having a similar impact on expelling the gas out. At a given circular velocity/stellar mass, this outflow possesses a velocity comparable to those fastest outflows in star-forming galaxies and shows a mass loading factor close to the highest values seen in the latter. However, the momentum and kinetic energy outflow rates of this outflow are close to the lowest values observed in the star-forming galaxies, which is likely caused by the 1--2 dex higher SFR in sources from \citet{Xu2022} than that in J0954$+$47. These results are all based on one dwarf galaxy with AGN, J0954$+$47, and a larger sample is needed before any robust conclusion can be drawn.}

{
\item 
The AGN in J0954$+$47 is energetic enough to easily drive the fast and powerful outflow in J0954$+$47. Nevertheless, we cannot formally rule out the possibility that star formation activity may also launch the outflow, assuming the momentum and energy provided by the stellar processes are transferred to the outflow at high efficiencies ($\gtrsim$60\%--230\% for momentum and $\gtrsim$30\%--120\% for kinetic energy), larger than typically observed in z$\sim$0 star-forming dwarf galaxies.}

\item 
The outflow in J0954$+$47 carries a significant amount of mass, momentum, and kinetic energy. With a mass loading factor (i.e., mass outflow rate over SFR) of $\gtrsim$3.4 ($\gtrsim$1.8) averaged over the entire outflow (for the high-v outflow), the outflow can likely transport more material out of the galaxy than that ends up in stars.
The kinetic energy outflow rate to AGN luminosity, \dedtlagn, is $\sim$0.6\% ($\sim$1.2\%) averaged over the entire outflow (for the high-v outflow) and is comparable to (larger than) the expectation from one representative simulation of AGN feedback. Additionally, this \dedt\ is $\sim$10$\times$ higher than that derived from the \oiiitext\ emission line from \citet{Liu2020}.

\item

A strong \heii\ emission line is detected in both J0954$+$47 and J1009$+$26 where the transition is covered by the observation. The EW and FWHM of the \heiitext, and the \civtext/\heiitext\ ratios are consistent with an AGN origin. Such dwarf AGN, if present in abundance in the early universe, may contribute to cosmic reionization. 

\item

In one object, J1009$+$26, the \heii\ and \oivb\ emission lines are the most blueshifted lines among all strong FUV and optical emission lines with respect to the systemic velocity in this object. These two emission lines likely trace a highly-ionized AGN wind in this object.

\end{itemize}

\clearpage

\appendix

\section{AGN Luminosity and SFR of J0954$+$47} \label{A1}

The AGN luminosity and SFR of J0954$+$47 are the two fundamental physical properties on which the discussions on the energy source and impact of the outflow are based. The AGN luminosity can also be obtained by integrating the AGN spectrum (scaled by the observed monochromatic continuum luminosity at 1320 \AA) adopted for the photoionization modeling in Section \ref{05321}, which gives log(\lagn) $\simeq$ 42.8 erg s$^{-1}$. This is an order of magnitude smaller than the \oiii\ based \lagn. Each of the two estimates of \lagn\ has its own caveat. For the FUV-based one, zero extinction is assumed and log(\lagn) may be underestimated. The departure of the intrinsic AGN radiation from the adopted AGN spectrum is also not known. For the \oiiitext-based one, the uncertainty of the scaling factor is mainly due to the large scatter of the scaling relation \citet[][]{Lamastra2009}. To make our analysis more self-consistent, we adopt the FUV-based \lagn\ in our following discussion on the outflow detected through the FUV absorption features.

As for the SFR, the \oiitext-based upper limit ($<$0.3 \msunyr), as reported in \citet{Liu2020}, is obtained by adopting the empirical relation calibrated for AGN only \citet{Ho2005}. In addition, two more upper limits, based on the FUV luminosity ($<$0.08 \msunyr) and FUV$+$MIR luminosities ($<$1.9 \msunyr), are reported in \citet{Latimer2021}. Both values are deemed upper limits as the scaling relations adopted are calibrated based on star-forming galaxies and the AGN contamination to the FUV and MIR luminosities are not subtracted. The FUV-only one may be a bit low as it does not account for the dust-obscured star formation activity. The FUV$+$MIR-based one is too high: J0954$+$47 is classified as AGN in all popular \textit{WISE}-based MIR diagnosis \citep[e.g.][]{Jarrett2011,Mateos2012}\footnote{As far as we are concerned, the only exception is the diagnosis proposed by \citet{Assef2010} adopting all 4 \textit{WISE} band.}. The MIR luminosity is thus likely dominated by the AGN rather than the star formation, leading to a significant overestimate of SFR. In all, we adopt the \oiitext-based upper limit on SFR in the following discussions.

\begin{figure}[!htb]   
\plotone{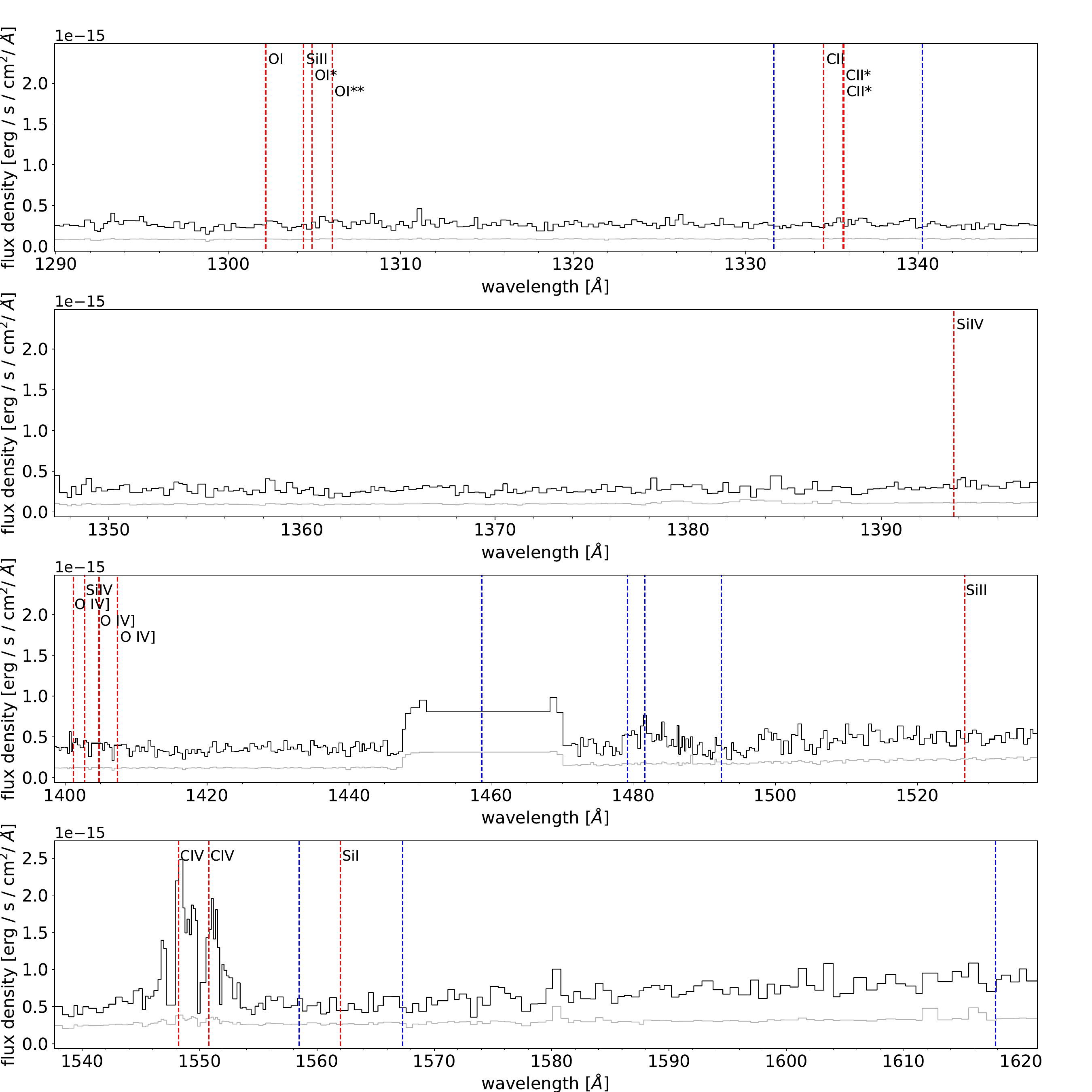}
\caption{Full G160M spectrum for J0906$+$56. The expected locations of strong transitions at the systemic velocity are marked in red dashed lines. The expected locations of strong z$=$0 foreground absorption features are marked in blue dashed lines.}
\label{fig:J0906all}
\end{figure}

\begin{figure}[!htb]   
\plotone{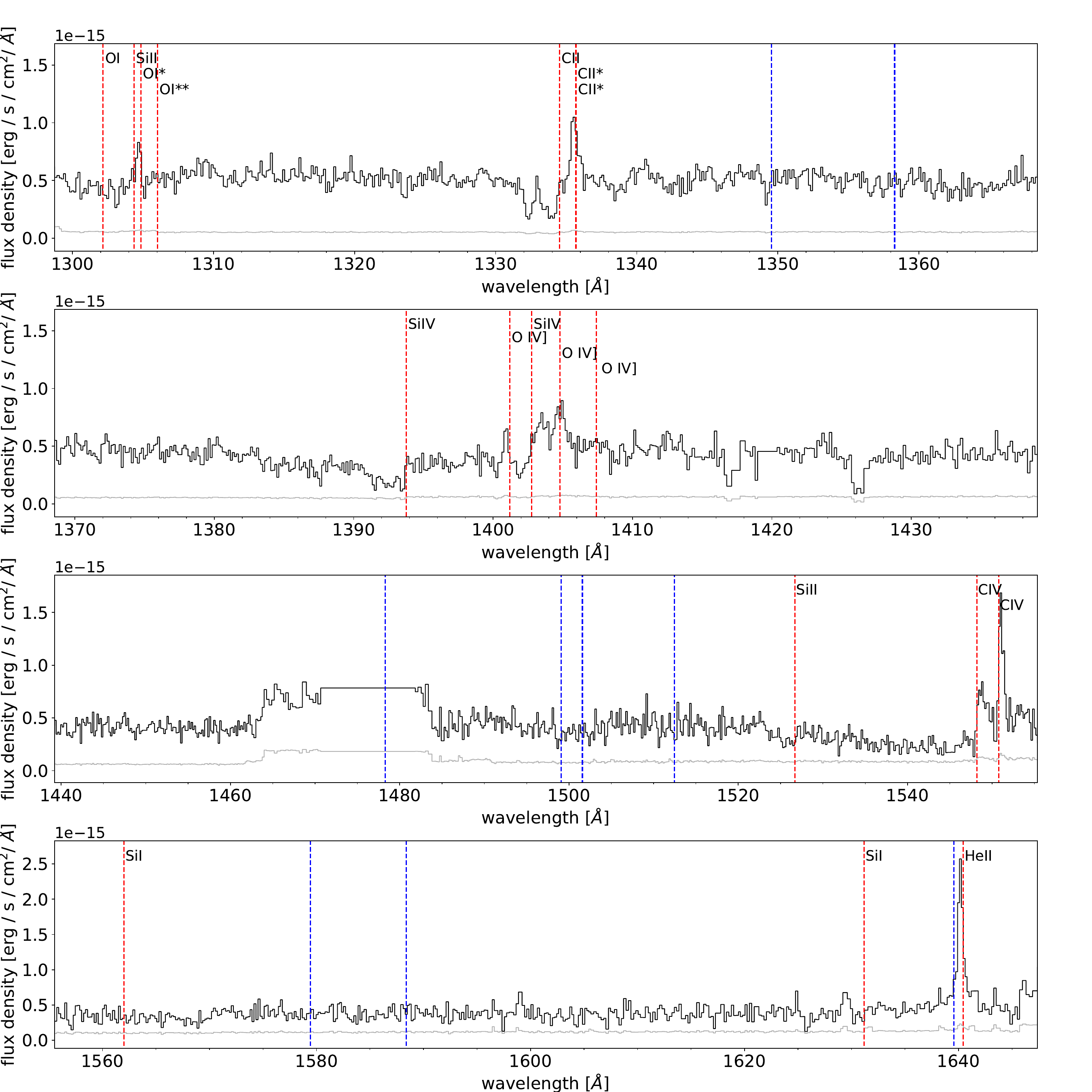}
\caption{Same as Fig. \ref{fig:J0906all} but for J0954$+$47.}
\label{fig:J0954all}
\end{figure}

\begin{figure}[!htb]   
\plotone{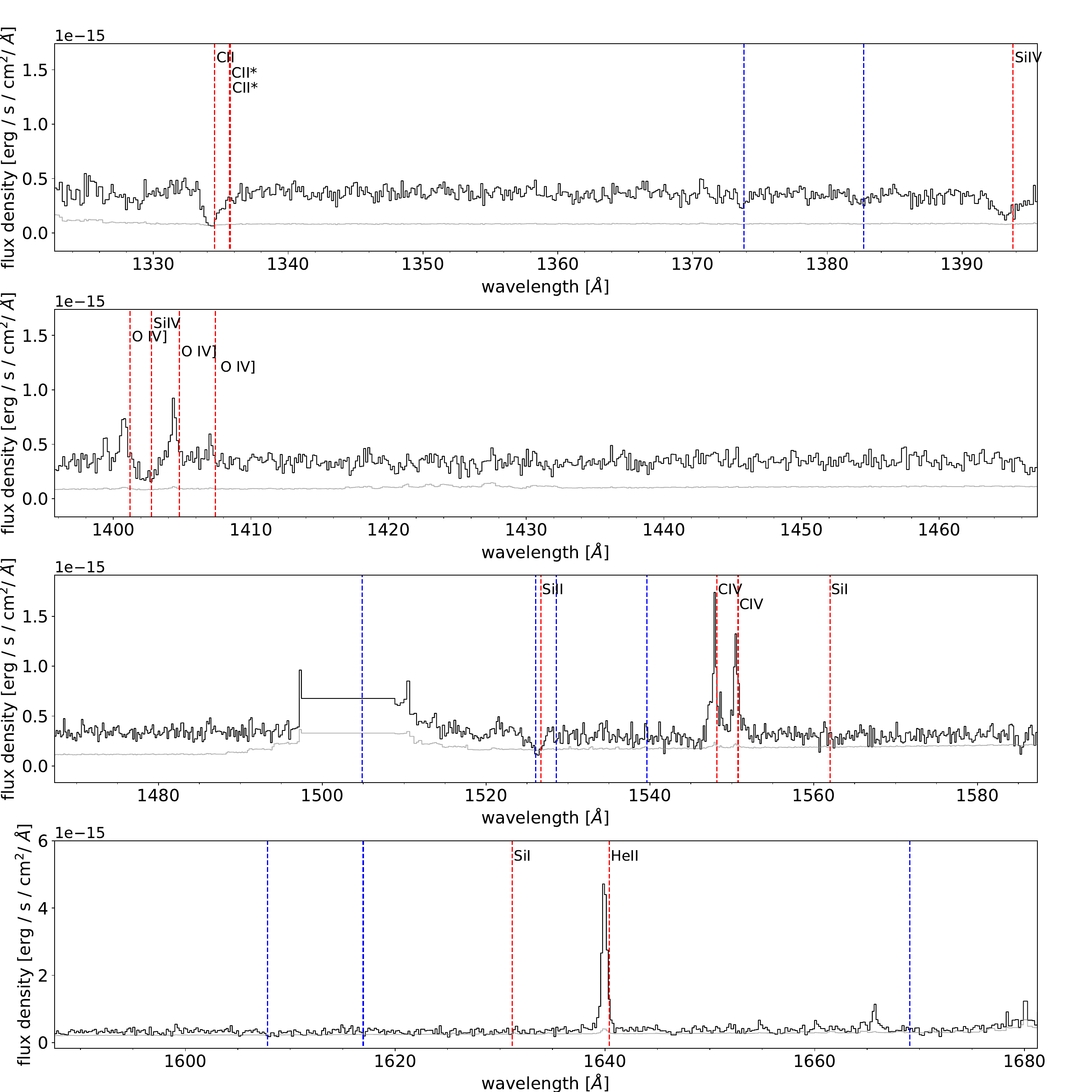}
\caption{Same as Fig. \ref{fig:J0906all} but for J1009$+$26.}
\label{fig:J1009all}
\end{figure}

\section{Full G160M Spectra of Our Objects} \label{A2}

The full \textit{HST}/COS G160M spectra of J0906$+$56, J0954$+$47 and J1009$+$26 are shown in Fig. \ref{fig:J0906all}, \ref{fig:J0954all} and \ref{fig:J1009all}.

\begin{acknowledgements} 
We thank the anonymous referee for constructive comments that have improved the paper. W.L. and S.V. acknowledge partial support for this work provided by NASA through grant numbers \textit{HST} GO-15662.001-A, GO-15662.001-B and 15915.002-A. T.M.T. acknowledges partial support for this work provided by NASA through grant numbers \textit{HST} GO-15915.004-A from the Space Telescope Science Institute, which is operated by AURA, Inc., under NASA contract NAS 5-26555. Based on observations made with the NASA/ESA Hubble Space Telescope, and obtained from the Hubble Legacy Archive, which is a collaboration between the Space Telescope Science Institute (STScI/NASA), the Space Telescope European Coordinating Facility (STECF/ESA) and the Canadian Astronomy Data Centre (CADC/NRC/CSA). This research has made use of the NASA/IPAC Extragalactic Database (NED), which is operated by the Jet Propulsion Laboratory, California Institute of Technology, under contract with the National Aeronautics and Space Administration.
\end{acknowledgements}

\facilities{HST(COS), HST(WFC3)}

\software{Astropy \citep{astropy2013, astropy2018}, BayesVP \citep{bvpfit}, CALCOS (\url{https://github.com/spacetelescope/calcos}),  ChiantiPy \citep{Dere1997,Dere2019}, CLOUDY \citep{CLOUDY}, LMFIT \citep{lmfit}, NumPy \citep{numpy}, SciPy \citep{scipy}.}


\bibliography{COSdwarf}{}
\bibliographystyle{aasjournal}



\end{document}